\documentclass[english,letter,amsmath,amssymb,showkeys,showpacs,aps]{revtex4}
\usepackage[T1]{fontenc}
\usepackage[latin9]{inputenc}
\usepackage{amsmath}
\usepackage{amssymb}
\usepackage{graphicx}
\usepackage{esint}

\makeatletter

\providecommand{\tabularnewline}{\\}

\@ifundefined{textcolor}{}
{%
 \definecolor{BLACK}{gray}{0}
 \definecolor{WHITE}{gray}{1}
 \definecolor{RED}{rgb}{1,0,0}
 \definecolor{GREEN}{rgb}{0,1,0}
 \definecolor{BLUE}{rgb}{0,0,1}
 \definecolor{CYAN}{cmyk}{1,0,0,0}
 \definecolor{MAGENTA}{cmyk}{0,1,0,0}
 \definecolor{YELLOW}{cmyk}{0,0,1,0}
}

\makeatother

\usepackage{babel}
\begin{document}

\title{Two Photon Exchange for Exclusive Pion Electroproduction}

\author{Andrei Afanasev}

\email{afanas@gwu.edu}

\affiliation{The George Washington University, Washington, DC 20052, USA}

\author{Aleksandrs Aleksejevs}

\email{aaleksejevs@swgc.mun.ca}

\affiliation{Memorial University of Newfoundland, Corner Brook, NL A2H 6P9, Canada}

\author{Svetlana Barkanova}

\email{svetlana.barkanova@acadiau.ca}

\affiliation{Acadia University, Wolfville, NS B4P 2R6, Canada}
\begin{abstract}
We perform detailed calculations of two-photon-exchange QED corrections to the cross section of pion electroproduction. The results are obtained with and without the soft-photon approximation; analytic expressions for the radiative corrections are derived. The relative importance of the two-photon correction is analyzed for the kinematics of several experiments at Jefferson Lab. A significant, over 20\%, effect due to two-photon exchange is predicted for the backward angles of electron scattering at large transferred momenta.
\end{abstract}


\maketitle

\section{Introduction}

The physics of hadrons, studied with electron scattering, is an active research field: Jefferson Lab is completing its successful
program using a 6-GeV CEBAF accelerator and is preparing a 12-GeV upgrade to start a new physics program 
within the next several years \citep{12GeV}.
Electron-scattering experiments probing the structure of hadrons also continue at the MAMI and ELSA facilities in Germany. 
A common feature of the modern experiments that study electromagnetic structure of hadrons is their high precision.
 
The increased accuracy of the measurements put stringent requirements on QED radiative corrections that must be applied to the data.
The largest QED corrections arise from bremsstrahlung, $i.e.$ the emission of real photons by the electrons. The bremsstrahlung corrections 
are characterized by large logarithms of the type $\log Q^2/m_e^2$, where $Q$ is transferred 4-momentum and $m_e$ is electron mass, 
resulting in an order-of-magnitude enhancement of $O(\alpha_{em})$ effects. 
Systematic uncertainties due to the bremsstrahlung corrections can be minimized by using iterative procedures and without requiring additional
knowledge of hadronic structure. However, when the experiments reach accuracy at a percent level, their interpretation also becomes sensitive to box-type contributions described by two-photon (2$\gamma$) exchange. 
A good example of the importance of $2\gamma$-exchange corrections is the Rosenbluth separation experiments in elastic electron-proton scattering \citep{TPE}. In these experiments, the measured Rosenbluth slope may be misinterpreted due to the omission of the angular dependence of $2\gamma$ corrections in the data analysis, which can mimic a significant part of the electric form factor contribution to the cross section.

In this paper, we evaluate  $2\gamma$-exchange effects for the processes of exclusive pion electroproduction
$e+p\,\rightarrow\, e+n+\pi^{+}$ and $e+p\,\rightarrow\, e+p+\pi^{0}$.
These processes have been studied experimentally in great detail and have become a valuable source
of information on the electromagnetic structure of nucleons. For example, depending on the choice of kinematics,
they provide access to the electromagnetic transition amplitudes from the nucleon ground state to excited states \citep{CLAS2008}; or $Q^2$-dependence of the pion electromagnetic form factor \citep{Garth1_2008, Garth2_2008, E07007, P1206101}; or, in a deep-virtual regime,  polarization-dependent Generalized Parton Distributions of a nucleon \citep{Ji97, Col97}.

The QED radiative corrections are indispensable in the interpretation of the experimental data on electron scattering.  
Although significant theoretical effort has been dedicated to this
problem, more work still needs to be done. A classical approach developed
by \citep{Mo&Tsai} four decades ago was mainly intended for inclusive
and elastic electron scattering, assuming integration over the phase space of final hadrons. More recent work addressed the case of electroproduction with detection of
hadrons in the final state for semi-inclusive \citep{Haprad} and exclusive diffractive \citep{Diffrad} processes. A calculation of QED corrections to exclusive
electroproduction of pions by \citep{Afan2002} provided
the next-to-leading order QED radiative corrections to the cross section
and the beam polarization asymmetry for exclusive pion electroproduction.
In comparison to earlier work  \citep{Mo&Tsai}, the authors of Ref.\citep{Afan2002} (a) eliminated dependence on a nonphysical parameter coming from
splitting soft and hard regions of the phase space of radiated photons
by covariant treatment of an infrared divergence using the
method of \citep{Bar77}, (b) treated bremsstrahlung exactly, without applying either the soft photon or peaking approximations and (c) retained the effects arising from the kinematics of detection in coincidence of two particles, an electron and a hadron, in the final state.
However, the calculation of \citep{Afan2002} only included vacuum
polarization corrections and contributions from bremsstrahlung and
vertex corrections from the electron. No box diagrams with two-photon exchange or photon
emission from hadrons were considered. 

Motivated by the need to evaluate two-photon exchange for measurements of the pion form factor, the authors
of Ref. \citep{Blunden_pi_2010} extended their
approach to elastic electron-proton scattering (outlined in \citep{Blunden_ep_2005})
to calculate two-photon exchange corrections 
in the elastic electron-pion scattering, including only pion elastic
intermediate states. The authors of Ref. \citep{Dong_and_Wong_2010} considered a process
of $e\pi^{+}$ elastic scattering as well. Both these papers, \citep{Blunden_pi_2010}
and \citep{Dong_and_Wong_2010}, concluded that the two-photon exchange
plays a sizable role in the pion form factor measurements at extreme backward angles.
In \citep{Borisyuk_pi_2010}, the two-photon exchange amplitude for
the elastic electron-pion scattering was computed in the dispersion
approach and with the monopole parameterization of the pion form factor,
including both elastic and inelastic contributions. 
All the previous calculations of 2$\gamma$-exchange for pion form factor measurements treated
the pion as a stable and free particle at rest, which complicates the application of this correction to actual measurements.

In the work presented here, we calculate the two-photon box corrections
for the cross sections of the entire process of pion electroproduction on a nucleon for two
production channels,

\begin{eqnarray*}
e\left(k_{1}\right)+p\left(k_{2}\right) & \rightarrow & e\left(k_{3}\right)+\pi^{+}\left(k_{4}\right)+n\left(k_{5}\right)\\
e\left(k_{1}\right)+p\left(k_{2}\right) & \rightarrow & e\left(k_{3}\right)+p\left(k_{4}\right)+\pi^{0}\left(k_{5}\right)
\end{eqnarray*}
where in the final state an electron and a charged hadron are detected, $\pi^+$ or the proton.
The calculations are performed both analytically and numerically using
\emph{FeynArts} and\emph{ FormCalc} \citep{LoopTools} as the base languages.
The regularization of infrared divergences arising in the box diagrams
is addressed by giving the photon a small rest mass and canceling
that by adding the soft-photon bremsstrahlung contribution from the electron,
proton, and the charged pion. 

The paper is constructed as follows: In Section II, we derive formulae for 2$\gamma$ corrections to pion electroproduction amplitudes. 

Section III describes the soft-photon bremsstrahlung contribution required for cancellation of infrared divergences. Section IV provides the analytical results.
Numerical analysis is performed in Section V, and the conclusions are given in Section V.

\section{Two Photon Box Diagrams in Pion Electroproduction}

Our goal is to derive the radiative correction associated with the two-photon box diagrams in exclusive
pion electroproduction. Here, we choose to employ, compare and analyze
two approaches. The first approach is based on the soft-photon approximation,
where we follow two prescriptions. In the first  prescription, we apply the soft-photon
approximation in the box amplitude consistently in the numerator and
denominator algebra according to the method suggested by Tsai \citep{Tsai}.
In this paper, we will call this type of soft-photon treatment
of the box amplitude SPT (Soft-Photon-Tsai). The second prescription
is based on the work of Maximon and Tjon \citep{Tjon}, where the soft-photon
approximation is applied only in the numerator algebra. Accordingly,
we will refer to this prescription as SPMT (Soft-Photon-Maximon-Tjon).
In the second approach, we evaluate the box amplitude exactly, without relying on the soft-photon approximation. Since in this case
the box amplitude can not be factorized by the Born amplitude, this approach
will result in the model dependence of the radiative correction. Here,
we evaluate the box amplitude by employing the model with the
monopole form-factor in the photon-proton, photon-pion and pion photoproduction
couplings. We will refer to this approach as FM (Formfactor-Model).

\begin{figure*}
\begin{centering}
\includegraphics[scale=0.9]{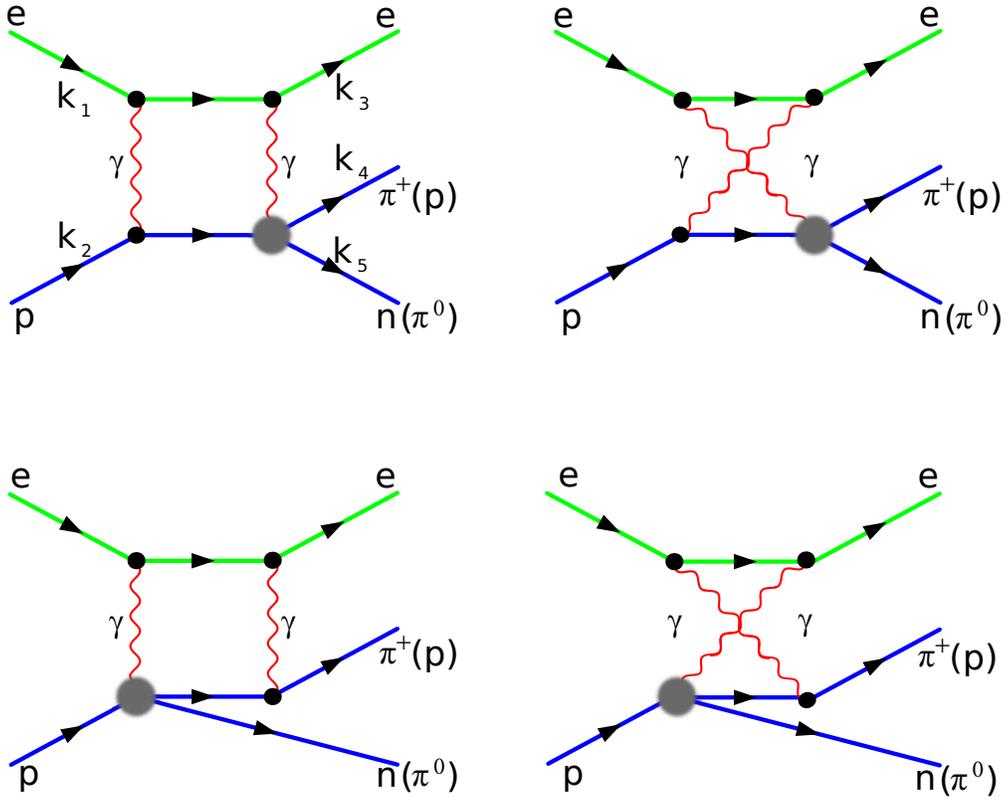}
\par\end{centering}

\caption{The two-photon box diagrams in pion electroproduction.}
\label{Flo:feynman_diag}
\end{figure*}

The two-photon box diagrams in exclusive pion electroproduction
are given in Fig.(\ref{Flo:feynman_diag}). For the incoming electron
and proton, we have momenta $k_{1}$ and $k_{2}$. For the outgoing
electron, detected proton (or pion: $\pi^{+}$) and undetected neutron
(or pion: $\pi^{0}$) we have momenta $k_{3},\, k_{4}$ and $k_{5},$
respectively.

Let us start with the definitions of the input kinematic parameters. We
choose to use the incoming electron energy in the laboratory frame $E_{lab}$,
the momentum transfer $Q^{2}=-q^{2}=-(k_{3}-k_{1})^{2}$ and the invariant
mass of the virtual photon plus incoming proton $W^{2}=\left(q+k_{2}\right)^{2}$
as input parameters. In addition, the polar ($\theta_{4}$) and azimuthal
($\phi_{4}$) angles of the detected proton (pion) are defined in the
center of mass of the final hadrons. The extended set of Lorentz invariants
is defined according to the prescription of \citep{Afan2002}. 
\begin{eqnarray}
 &  & S=2\left(k_{1}k_{2}\right)=2m_{2}E_{lab},\nonumber \\
\nonumber \\
 &  & {\displaystyle X=2\left(k_{3}k_{2}\right)=S-W^{2}-Q^{2}+m_{2}^{2},}\nonumber \\
\nonumber \\
 &  & S_{t}=2\left(k_{2}k_{4}\right)=W^{2}+m_{4}^{2}-m_{5}^{2}+V_{3}-V_{1},\label{eq:kin1}\\
\nonumber \\
 &  & {\displaystyle V_{1,3}=2\left(k_{1,3}k_{4}\right)=2\left(E_{1,3}E_{4}-p_{1,3}p_{4}\left(\cos\theta_{4}\cos\theta_{1,3}+\sin\theta_{4}\sin\theta_{1,3}\cos\phi_{4}\right)\right)},\nonumber \\
\nonumber \\
 &  & u_{1}=S-Q^{2},\, u_{3}=X+Q^{2},\nonumber \\
\nonumber \\
 &  & \lambda_{1,3}=u_{1,3}^{2}-4m_{1}^{2}W^{2},\,\,\,\,\lambda_{q}=\left(S-X\right)^{2}+4m_{2}^{2}Q^{2}.\nonumber 
\end{eqnarray}
With the help of the invariants in Eq.(\ref{eq:kin1}), we define energy and momenta in the center of mass system of the virtual
photon $q=k_{1}-k_{3}$ and initial proton as follows: 
\begin{flalign}
 & E_{1,3}=\frac{u_{1,3}}{2W},\,\,\,\,\, p_{1,3}=\frac{\sqrt{\lambda_{1,3}}}{2W},\nonumber \\
\nonumber \\
 & E_{2}=\frac{S-X+2m_{2}^{2}}{2W},\,\,\,\,\, p_{2}=p_{q}=\frac{\sqrt{\lambda_{q}}}{2W},\label{eq:kin2}\\
\nonumber \\
 & E_{4}=\frac{W^{2}+m_{4}^{2}-m_{5}^{2}}{2W},\,\,\,\,\, p_{4}=\frac{\sqrt{\left(W^{2}+m_{4}^{2}-m_{5}^{2}\right)^{2}-4m_{4}^{2}W^{2}}}{2W}.\nonumber 
\end{flalign}
The angles $\theta_{1,3}$ are given by the following expressions:
\begin{flalign}
 & \cos\theta_{1,3}=\frac{u_{1,3}\left(S-X-2Q^{2}\right)\pm2Q^{2}W^{2}}{\sqrt{\lambda_{1,3}}\sqrt{\lambda_{q}}},\label{eq:kin3}\\
\nonumber \\
 & \sin\theta_{1,3}=\frac{2W\sqrt{\lambda}}{\sqrt{\lambda_{1,3}}\sqrt{\lambda_{q}}},\,\,\,\,\,\lambda=Q^{2}u_{1}u_{2}-Q^{4}W^{2}-m_{1}^{2}\lambda_{q}.\nonumber 
\end{flalign}

\subsection{Soft-photon Approximation}

The soft-photon approximation significantly simplifies the calculations of the two-photon box amplitudes. Specifically,
a photon which does not couple to the pion photoproduction vertex
is treated as a soft one. This implies that we extract only the infrared-divergent
part of the amplitude and, for now, we assume that the non-infrared-divergent
part of the amplitude gives a negligible contribution. The infrared
divergences arising in the cross section from the two-photon box diagrams will be treated later using soft-photon bremsstrahlung. 

As an example, let us start with the first graph on Fig.(\ref{Flo:feynman_diag}(a)).
For this graph, we can write the following:

\begin{eqnarray}
{\displaystyle M_{1}} & = & {\displaystyle \frac{i}{16\pi^{4}}\int d^{4}q\left[\bar{u}(k_{3},m_{3})\left(ie\gamma_{\mu}\right)\frac{\not k_{1}-\not q+m_{1}}{\left(k_{1}-q\right)^{2}-m_{1}^{2}}\left(ie\gamma_{\nu}\right)u(k_{1},m_{1})\right]\cdot}\nonumber \\
 &  & {\displaystyle \left[\bar{u}(k_{5(4)},m_{5(4)})\left(\Gamma_{\alpha}^{p-\gamma-\pi-n(p)}\right)\frac{\not k_{2}+\not q+m_{2}}{\left(k_{2}+q\right)^{2}-m_{2}^{2}}\left(\Gamma_{\beta}^{p-\gamma-p}(q)\right)u(k_{2},m_{2})\right]\cdot\frac{g^{\nu\beta}}{q^{2}}\cdot\frac{g^{\mu\alpha}}{(q+k_{3}-k_{1})^{2}}}.\label{eq:a1}
\end{eqnarray}
Here, $m_{5}=m_{n}$ for the case of $\pi^{+}$ production and $m_{4}=m_{p}$
for $\pi^{0}$ production. The photon-proton coupling has the
following structure: $\Gamma_{\beta}^{p-\gamma-p}(q)=-ie\gamma_{\beta}F(q)$,
with the form-factor $F(q)=\frac{\Lambda^{2}}{\Lambda^{2}-q^{2}}$
($\Lambda^{2}=0.83m_{p}^{2}$) taken in the monopole form. In the
case of the soft-photon approximation with either the SPT or SPMT ($q\rightarrow0\, GeV)$
prescriptions, the coupling $\Gamma_{\beta}^{p-\gamma-p}(q)$ converges
to the coupling of the point-like proton $\Gamma_{\beta}^{p-\gamma-p}(q\rightarrow0)=-ie\gamma_{\beta}$,
and the box amplitude becomes independent of the choice of the type
of form-factor. We do not show the explicit structure
of the pion photoproduction coupling $\Gamma_{\alpha}^{p-\gamma-\pi-n(p)}$ here because the structure of that coupling has no impact
on the two-photon box radiative correction calculated in the soft-photon approximation.

Using the soft-photon approximation according to the SPT prescription,
we neglect the momentum of the photon in the numerator algebra and in
the denominator of photon propagator: $\frac{g^{\mu\alpha}}{\left(q+k_{3}-k_{1}\right)^{2}}\rightarrow\frac{g^{\mu\alpha}}{\left(k_{3}-k_{1}\right)^{2}}$.
As a result, with the help of the Dirac equation, terms such as
$(\not k_{i}+m_{i})\left(ie\gamma_{\nu}\right)u(k_{i},\, m_{i})$
can be simplified into $2ie(k_{i})_{\nu}$. For the amplitude in Eq.(\ref{eq:a1})
we can write:
\begin{eqnarray}
{\displaystyle M_{1}^{SPT}} & = & {\displaystyle \frac{i\alpha}{4\pi^{3}}\left[\bar{u}(k_{3},m_{3})\left(ie\gamma_{\mu}\right)u(k_{1},m_{1})\right]\left[\bar{u}(k_{5(4)},m_{5(4)})\left(\Gamma_{\alpha}^{p-\gamma-\pi-n(p)}\right)u(k_{2},m_{2})\right]\cdot\frac{g^{\mu\alpha}}{(k_{3}-k_{1})^{2}}\cdot}\nonumber \\
 &  & {\displaystyle 4\left(k_{1}\cdot k_{2}\right)\int d^{4}q\frac{1}{q^{2}}\cdot\frac{1}{\left(k_{2}+q\right)^{2}-m_{2}^{2}}\cdot\frac{1}{\left(k_{1}-q\right)^{2}-m_{1}^{2}}.}\label{eq:a2}
\end{eqnarray}
Taking into account that the Born-level amplitude is
\begin{eqnarray}
M_{0} & = & \left[\bar{u}(k_{3},m_{3})\left(ie\gamma_{\mu}\right)u(k_{1},m_{1})\right]\left[\bar{u}(k_{5(4)},m_{5(4)})\left(\Gamma_{\alpha}^{p-\gamma-\pi-n(p)}\right)u(k_{2},m_{2})\right]\cdot\frac{g^{\mu\alpha}}{(k_{3}-k_{1})^{2}},\label{eq:a3}
\end{eqnarray}
we can write
\begin{eqnarray}
M_{1}^{SPT} & = & -\frac{\alpha}{2\pi}S\cdot C_{0}\left(\{k_{1},m_{1}\},\{-k_{2},m_{2}\}\right)\cdot M_{0}.\label{eq:a4}
\end{eqnarray}
Here, $C_{0}\left(\{k_{i},m_{i}\},\{k_{j},m_{j}\}\right)$ is the Passarino-Veltman
three-point scalar integral defined as:
\begin{eqnarray}
C_{0}\left(\{k_{i},m_{i}\},\{k_{j},m_{j}\}\right) & = & \frac{1}{i\pi^{2}}\int d^{4}q\frac{1}{q^{2}}\cdot\frac{1}{\left(k_{i}-q\right)^{2}-m_{i}^{2}}\cdot\frac{1}{\left(k_{j}-q\right)^{2}-m_{j}^{2}}.\label{eq:a5}
\end{eqnarray}
A similar approach can be employed for the rest of the graphs in
Fig.(\ref{Flo:feynman_diag}). The results for graphs
of types\textit{} (b), (c), and (d) can be written in the following form:
\begin{eqnarray}
M_{2}^{SPT} & = & -\frac{\alpha}{2\pi}X\cdot C_{0}\left(\{k_{3},m_{3}\},\{k_{2},m_{2}\}\right)\cdot M_{0},\label{eq:a6a}\\
M_{3}^{SPT} & = & -\frac{\alpha}{2\pi}V_{3}\cdot C_{0}\left(\{k_{3},m_{3}\},\{-k_{4},m_{4}\}\right)\cdot M_{0},\label{eq:a6b}\\
M_{4}^{SPT} & = & -\frac{\alpha}{2\pi}V_{1}\cdot C_{0}\left(\{k_{1},m_{1}\},\{k_{4},m_{4}\}\right)\cdot M_{0}.\label{eq:a6c}
\end{eqnarray}
Combining Eqs.(\ref{eq:a4}), (\ref{eq:a6a}), (\ref{eq:a6b}) and (\ref{eq:a6c})
we can write the total amplitude for the two-photon box diagrams in
the following form
\begin{eqnarray}
M_{box}^{SPT}=\sum_{j=1}^{4}M_{j}^{I}=-\frac{\alpha}{2\pi}\left(S\cdot C_{0}\left(\{k_{1},m_{1}\},\{-k_{2},m_{2}\}\right)+X\cdot C_{0}\left(\{k_{3},m_{3}\},\{k_{2},m_{2}\}\right)+\right.\nonumber \\
\left.V_{3}\cdot C_{0}\left(\{k_{3},m_{3}\},\{-k_{4},m_{4}\}\right)+V_{1}\cdot C_{0}\left(\{k_{1},m_{1}\},\{k_{4},m_{4}\}\right)\right)\cdot M_{0}.
\end{eqnarray}
 It is worth mentioning that in the case where $\pi^{+}$ appears
in the loop (Fig.\ref{Flo:feynman_diag} (c, d)) of the box diagram,
we use the $\pi-\gamma-\pi$ coupling in the following form:
$\Gamma_{\alpha}^{\pi-\gamma-\pi}=-ie\cdot(k_{\pi}^{in}+k_{\pi}^{out})_{\alpha}F(q)$.
The pion propagator is given by $\Pi_{\pi}=\frac{i}{k_{\pi}^{2}-m_{\pi}^{2}}$.
When the soft-photon approximation is applied, the photon-pion coupling
becomes $\Gamma_{\alpha}^{\pi-\gamma-\pi}\rightarrow-2ie\cdot(k_{4})_{\alpha}$
and hence the box amplitude will have the same general structure as
for the case of $\pi^{0}$ electroproduction. 

It is obvious that in Eqs.(\ref{eq:a4}) and (\ref{eq:a6a}-\ref{eq:a6c}),
the box amplitude is factorized by the Born amplitude, which can be evaluated in the different models or approaches. Keeping
terms of order $\mathcal{O}(\alpha^{3})$ only, the factorization
of the pion-electroproduction cross section can be accomplished in
the following way:
\begin{eqnarray}
d\sigma^{SPT}=\left(|M_{0}|^{2}+2\mbox{Re}\left[M_{box}^{SPT}M_{0}^{\dagger}\right]+\mathcal{O}\left(\alpha^{4}\right)\right)d\Gamma_{2\rightarrow3}=d\sigma_{0}+d\sigma_{box}^{SPT}=d\sigma_{0}\cdot\left(1+\delta_{box}^{SPT}\right),\label{eq:a8}
\end{eqnarray}
where $d\sigma_{0}$ is the Born differential cross section, $d\Gamma_{2\rightarrow3}$
is the phase factor for the $2\rightarrow3$ process and $\delta_{box}^{SPT}=\frac{d\sigma_{box}^{SPT}}{d\sigma_{0}}$
is the radiative correction due to the box diagrams contribution:
\begin{eqnarray}
\delta_{box}^{SPT}=-\frac{\alpha}{\pi}\mbox{Re}\big[S\cdot C_{0}\left(\{k_{1},m_{1}\},\{-k_{2},m_{2}\}\right)+X\cdot C_{0}\left(\{k_{3},m_{3}\},\{k_{2},m_{2}\}\right)+\nonumber \\
V_{3}\cdot C_{0}\left(\{k_{3},m_{3}\},\{-k_{4},m_{4}\}\right)+V_{1}\cdot C_{0}\left(\{k_{1},m_{1}\},\{k_{4},m_{4}\}\right)\big].\label{eq:a9}
\end{eqnarray}
Since the dependence on the pion photoproduction coupling
$\Gamma_{\alpha}^{p-\gamma-\pi-n(p)}$ is canceled in the soft two-photon
box radiative correction, we can call this type of calculation a model-independent type.

The three-point scalar integral $C_{0}\left(\{k_{i},m_{i}\},\{k_{j},m_{j}\}\right)$
in Eq.(\ref{eq:a9}) was evaluated both using an approximated expression
($m_{i}\ll m_{j}$) taken from \citep{Kuraev2006}, and exactly with
the help of the package LoopTools \citep{LoopTools}. The explicit expression
for this integral reads as follows:
\begin{eqnarray}
{\displaystyle C_{0}\left(\{k_{i},m_{i}\},\{k_{j},m_{j}\}\right)=-\frac{1}{4\left(k_{i}\cdot k_{j}\right)}\left[\ln^{2}\frac{2\left(k_{i}\cdot k_{j}\right)}{m_{j}^{2}}+\left(2\ln\frac{2\left(k_{i}\cdot k_{j}\right)}{m_{j}^{2}}+\ln\frac{m_{j}^{2}}{m_{i}^{2}}\right)\ln\frac{m_{j}^{2}}{\lambda^{2}}-\right.}\nonumber \\
\left.\frac{1}{2}\ln^{2}\frac{m_{j}^{2}}{m_{i}^{2}}-2Li_{2}\left(-\frac{m_{j}^{2}-2\left(k_{i}\cdot k_{j}\right)}{2\left(k_{i}\cdot k_{j}\right)}\right)\right],\label{eq:a10a}
\end{eqnarray}
where $Li_{2}(z)=-\int_{0}^{z}\frac{\ln(1-t)}{t}dt$ is the usual dilogarithm
function. It is important to note that in our calculations we did not apply
the approximation $\mbox{Re}\left[C_{0}\left(\{k_{i},m_{i}\},\{-k_{j},m_{j}\}\right)\right]\thickapprox-C_{0}\left(\{k_{i},m_{i}\},\{k_{j},m_{j}\}\right)$
introduced earlier in \citep{Tsai}. As a result, this produced a shift
of $\alpha\pi$ in the value of the box radiative correction using
the SPT prescription.

When evaluating the two-photon box amplitude using the SPMT prescription
of the soft-photon approximation, we follow the same steps as in
SPT, but we no longer neglect the photon momentum $(q)$ in the
denominator of the photon propagator $\frac{g^{\mu\alpha}}{(q+k_{3}-k_{1})^{2}}$.
This effectively results in the replacement of the three-point scalar
integrals by the four-point ones. Specifically, we can write: 
\begin{eqnarray}
\delta_{box}^{SPMT}=-\frac{\alpha Q^{2}}{\pi}\mbox{Re}\big[S\cdot D_{0}\left(\{k_{1},m_{1}\},\{-k_{2},m_{2}\},\{k_{1},k_{3}\}\right)+X\cdot D_{0}\left(\{k_{3},m_{3}\},\{k_{2},m_{2}\},\{-k_{1},-k_{3}\}\right)+\nonumber \\
+V_{3}\cdot D_{0}\left(\{k_{3},m_{3}\},\{-k_{4},m_{4}\},\{-k_{1},-k_{3}\}\right)+V_{1}\cdot D_{0}\left(\{k_{1},m_{1}\},\{k_{4},m_{4}\},\{k_{1},k_{3}\}\right)\bigr].\label{eq:a10b}
\end{eqnarray}
The four-point scalar integral $D_{0}$ has the following structure:
\begin{equation}
D_{0}\left(\{k_{i},m_{i}\},\{k_{j},m_{j}\},\{k_{\alpha},k_{\beta}\}\right)=\frac{1}{i\pi^{2}}\int d^{4}q\frac{1}{q^{2}}\cdot\frac{1}{\left(k_{i}-q\right)^{2}-m_{i}^{2}}\cdot\frac{1}{\left(k_{j}-q\right)^{2}-m_{j}^{2}}\cdot\frac{1}{(q+k_{\beta}-k_{\alpha})^{2}}\label{eq:a10c}
\end{equation}
and can be numerically evaluated using the package LoopTools. To see how
our results in Eq.(\ref{eq:a10c}) compare to the results of \citep{Tjon},
we have reduced Eq.(\ref{eq:a10b}) to the case of elastic electron-proton scattering and for that the corresponding correction can be written in the form:
\begin{eqnarray}
\delta_{box-(ep)}^{SPMT}=-\frac{\alpha Q^{2}}{\pi}\mbox{Re}\big[S\cdot D_{0}\left(\{k_{1},m_{1}\},\{-k_{2},m_{2}\},\{k_{1},k_{3}\}\right)+X\cdot D_{0}\left(\{k_{3},m_{3}\},\{k_{2},m_{2}\},\{-k_{1},-k_{3}\}\right)\big].\label{eq:a10d}\\
\nonumber 
\end{eqnarray}
Here, since this is elastic $e-p$ scattering, we take $m_{\pi}=0\,(GeV)$
and invariant mass $W=m_{p}$, which effectively reduces the kinematics
of the $2\rightarrow3$ process to those of a $2\rightarrow2$ process. After the treatment of
infrared divergences in Eq.(\ref{eq:a10d}) with soft-photon bremsstrahlung,
we find that the results produced in Eq.(11) from \citep{Tjon} (for
the case of $\gamma-\gamma$ box only) and Eq.(\ref{eq:a10d}) are
identical.
Such an approach was considered by Maximon and Tjon \citep{Tjon} as a "less drastic" version of the soft-photon approximation. We note, however, that
since it treats numerators and denominators of fermion propagators differently, it violates Ward identity, i.e. it does not preserve the gauge invariance. In principle, such an approximation
would lead to the non-renormalizability of the theory. Luckily, for the specific case of the two-photon exchange calculations, it does not result in any unphysical divergences.

\subsection{Exact Model-dependent Approach}

In the second approach, we apply the computational methods described
in \citep{LoopTools} and calculate the two-photon box correction without
relying on the soft-photon approximation. In this case, the two-photon
box amplitude is calculated exactly, but since it is not possible
to factorize it by Born amplitude, this approach becomes effectively
model-dependent. Here, we define a radiative correction using the Formfactor-Model
(FM) notation in the following form:

\begin{eqnarray}
\delta_{box}^{FM}=\frac{d\sigma_{box}^{FM}}{d\sigma_{0}}=\frac{2\mbox{Re}\left[M_{box}^{FM}M_{0}^{\dagger}\right]}{|M_{0}|^{2}}.\label{eq:a10e}
\end{eqnarray}
In order to calculate the radiative correction $\delta_{box}^{FM}$ to
the Born cross section, we employ a model where the monopole form-factor
is applied in every photon-hadron coupling. In this case, the pion
photoproduction coupling is described by the axial-vector contact
term $\Gamma_{\alpha}^{p-\gamma-\pi-n(p)}=\frac{e(D+F)}{f_{\pi}}\gamma_{\alpha}\gamma_{5}F(q)$,
which is weighted by the form-factor taken in monopole form $F(q)=\frac{\Lambda^{2}}{\Lambda^{2}-q^{2}}$.
The low-energy constants $D=0.40\pm0.03$ and $F=0.61\pm0.04$ were
determined earlier in \citep{Manohar} and $f_{\pi}=\sqrt{2}\cdot92\, MeV$
is the pion coupling constant. The photon-proton and photon-pion couplings
also have the same structure as in the case of SPT or SPMT: $\Gamma_{\beta}^{p-\gamma-p}(q)=-ie\gamma_{\beta}F(q)$
and $\Gamma_{\alpha}^{\pi-\gamma-\pi}=-ie\cdot(k_{\pi}^{in}+k_{\pi}^{out})_{\alpha}F(q)$, 
respectively. We do not neglect the momentum of
the photon in the form-factor when evaluating the loop integral. Calculations
of the two-photon box radiative correction where done analytically
using FeynArts and FormCalc and later numerically using LoopTools.
Although we have analytic expressions derived for the correction
in Eq.(\ref{eq:a10e}) as well, it is not possible to show them in
this paper due to their length. For this approach, we show numerical
analysis only, which is presented in Section V.

\section{Soft-Photon Bremsstrahlung}

\begin{figure*}
\centering{}\includegraphics[scale=0.6]{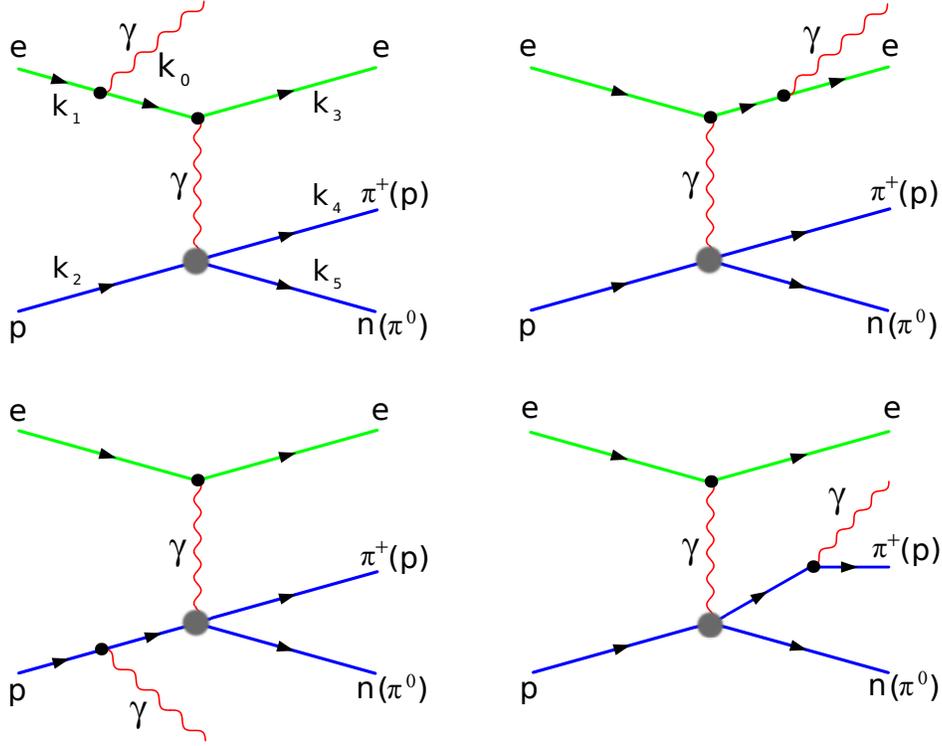}\caption{Representative graphs for the bremsstrahlung process in pion electroproduction.
Here, the momenta distribution is the same as in the case of box calculations
with momenta of the photon taken as $k_{0}$.}
\label{Flo:soft}
\end{figure*}

At this point we are ready to address the infrared divergences arising
in the box diagrams. Regularization of the infrared divergent integrals
in Eq.(\ref{eq:a5}) and (\ref{eq:a10c}) can be done by assigning
a small rest mass $\lambda$ to the photon. Clearly, the photon mass
should be removed to avoid nonphysical representation of the final
results. This can be achieved by adding the soft-photon bremsstrahlung
contribution coming from the graphs in Fig.(\ref{Flo:soft}).

As an example, we consider the first graph in Fig.(\ref{Flo:soft})
and evaluate the amplitude in the soft-photon emission approximation
by neglecting the momentum of the photon in the numerator algebra.
Hence, we can write:
\begin{eqnarray}
 & M_{1,\gamma}={\displaystyle i\left[\overline{u}\left(k_{3},m_{3}\right)\left(ie\gamma_{\mu}\right)u(k_{1},m_{1})\right]\cdot}\label{eq:a10}\\
 & {\displaystyle \left[\bar{u}(k_{5},m_{5})\left(\Gamma_{\nu}^{p-\gamma-\pi^{+}-n}\right)\frac{\not k_{2}-\not k_{0}+m_{2}}{\left(k_{2}-k_{0}\right)^{2}-m_{2}^{2}}\left(\Gamma_{\alpha}^{p-\gamma-p}(k_{0})\right)u(k_{2},m_{2})\right]\cdot\frac{g^{\mu\nu}}{\left(k_{3}-k_{1}\right)^{2}}\epsilon^{\star,\alpha}\left(k_{0}\right)}\nonumber 
\end{eqnarray}
Here, $\epsilon^{\star,\alpha}\left(k_{0}\right)$ is the polarization
vector of the emitted photon. In the limit where the energy of the emitted
photon is small, we can readily assume that $\Gamma_{\alpha}^{p-\gamma-p}(k_{0})\rightarrow\Gamma_{\alpha}^{p-\gamma-p}(0)=iQ_{p}\gamma_{\alpha}$
and $\frac{\not k_{2}-\not k_{0}+m_{2}}{\left(k_{2}-k_{0}\right)^{2}-m_{2}^{2}}\left(iQ_{p}\gamma_{\alpha}\right)u(k_{2},m_{2})$
can be reduced to $-iQ_{p}\frac{\left(k_{2}\right)_{\alpha}}{\left(k_{2}\cdot k_{0}\right)}u(k_{2},m_{2})$
with the help of the Dirac equation. As a result, we can write the following
for the amplitude in Eq.(\ref{eq:a10a}): 
\begin{eqnarray*}
 & {\displaystyle {\displaystyle M_{1,\gamma}=M_{0}\cdot Q_{p}\frac{\left(k_{2}\epsilon^{\star}\left(k_{0}\right)\right)}{\left(k_{2}\cdot k_{0}\right)}.}}
\end{eqnarray*}
We can write bremsstrahlung amplitudes for the rest
of the graphs in a similar way:
\begin{eqnarray}
 & {\displaystyle M_{1,\gamma}=M_{0}\cdot\left[e\frac{\left(k_{1}\epsilon^{\star}\left(k_{0}\right)\right)}{\left(k_{1}\cdot k_{0}\right)}+Q_{p}\frac{\left(k_{2}\epsilon^{\star}\left(k_{0}\right)\right)}{\left(k_{2}\cdot k_{0}\right)}-e\frac{\left(k_{3}\epsilon^{\star}\left(k_{0}\right)\right)}{\left(k_{3}\cdot k_{0}\right)}-Q_{\pi^{+}(p)}\frac{\left(k_{4}\epsilon^{\star}\left(k_{0}\right)\right)}{\left(k_{4}\cdot k_{0}\right)}\right].}\label{eq:a11}
\end{eqnarray}
It is obvious from the equation above that the cross section for the
bremsstrahlung process is parameterized by the Born cross section as:
\begin{eqnarray}
 & {\displaystyle d\sigma_{\gamma}=d\sigma_{0}\cdot\frac{1}{(2\pi)^{3}}\int\frac{d^{3}\mathbf{k}_{0}}{2\omega}} & {\displaystyle \left[e\frac{\left(k_{1}\epsilon^{\star}\left(k_{0}\right)\right)}{\left(k_{1}\cdot k_{0}\right)}+Q_{p}\frac{\left(k_{2}\epsilon^{\star}\left(k_{0}\right)\right)}{\left(k_{2}\cdot k_{0}\right)}-e\frac{\left(k_{3}\epsilon^{\star}\left(k_{0}\right)\right)}{\left(k_{3}\cdot k_{0}\right)}-Q_{\pi^{+}(p)}\frac{\left(k_{4}\epsilon^{\star}\left(k_{0}\right)\right)}{\left(k_{4}\cdot k_{0}\right)}\right]^{2}}\label{eq:a12}
\end{eqnarray}
Here, $\omega$ is the energy of the emitted photon. For the calculations
of the bremsstrahlung cross section, we have to account only for the
parts of Eq.(\ref{eq:a12}) which are responsible for the treatment
of infrared divergences in boxes. Namely, the products of graphs one and three, one and four, and two and three
plus two and four in Fig.(\ref{Flo:soft}). After summing over
all polarization states of the emitted photon and using $\sum_{\epsilon}\left(k_{i}\epsilon\left(k_{0}\right)\right)\left(k_{j}\epsilon^{\star}\left(k_{0}\right)\right)=-\left(k_{i}k_{j}\right)$,
we can write, for the part of the bremsstrahlung cross section responsible
for the IR-divergence cancellation in boxes, the following result: 
\begin{eqnarray}
 & {\displaystyle d\sigma'_{\gamma}=d\sigma_{0}\cdot\delta_{\gamma},}\label{eq:a12b}\\
 & {\displaystyle \delta_{\gamma}=-\frac{\alpha}{2\pi^{2}}\int\frac{d^{3}\mathbf{k}_{0}}{\omega}\left[-\frac{\left(k_{1}\cdot k_{2}\right)}{\left(k_{1}\cdot k_{0}\right)\left(k_{2}\cdot k_{0}\right)}+\frac{\left(k_{1}\cdot k_{4}\right)}{\left(k_{1}\cdot k_{0}\right)\left(k_{4}\cdot k_{0}\right)}+\frac{\left(k_{3}\cdot k_{2}\right)}{\left(k_{3}\cdot k_{0}\right)\left(k_{2}\cdot k_{0}\right)}-\frac{\left(k_{3}\cdot k_{4}\right)}{\left(k_{3}\cdot k_{0}\right)\left(k_{4}\cdot k_{0}\right)}\right],}\nonumber 
\end{eqnarray}
where $\delta_{\gamma}$ is defined as the soft-photon bremsstrahlung
correction. Here we have used $Q_{\pi^{+}(p)}=-e$. The integral in
Eq.(\ref{eq:a12b}) clearly has IR-divergent behavior ,and its  regularization
can be achieved in the same way as before by giving the photon
a fictitious mass $(\lambda)$ with the photon's energy being defined
as $\omega=\sqrt{\mathbf{k}_{0}^{2}+\lambda^{2}}$. Eq.(\ref{eq:a12b})
is presented in an invariant form, so integration over the phase
space of the emitted photon can be carried out in any reference frame.
We choose to follow the prescription of \citep{Tjon} where integration
is carried out in \textbf{R} frame defined as $\mathbf{k_{\mathbf{5}}=-k}_{0}$.
With the final results written in invariant form, we list only
the final equations relevant to this work. The soft-photon
integral can be written as follows:
\begin{eqnarray}
 & {\displaystyle I\left(k_{i},k_{j}\right)=\int\frac{d^{3}\mathbf{k}_{0}}{\omega}\frac{1}{\left(k_{i}\cdot k_{0}\right)\left(k_{j}\cdot k_{0}\right)}}\label{eq:a14}
\end{eqnarray}
The integral $I\left(k_{i},k_{j}\right)$ is evaluated as:
\begin{eqnarray}
 & {\displaystyle I\left(k_{i},k_{j}\right)=\frac{2\pi}{\gamma_{ij}}\cdot\left[A_{ij}^{(1)}+A_{ij}^{(2)}\right]}\label{eq:a15}
\end{eqnarray}
where for $i\neq j$ we get
\begin{eqnarray}
 & {\displaystyle A_{ij}^{(1)}=\ln\left(\alpha_{ij}\frac{m_{i}}{m_{j}}\right)\cdot\ln\left(\frac{4{\displaystyle \Delta\varepsilon^{2}}}{\lambda^{2}}\right)}\label{eq:a16}
\end{eqnarray}
and
\begin{eqnarray}
 & {\displaystyle A_{ij}^{(2)}=\ln^{2}\left(\frac{\beta_{i}}{m_{i}\sqrt{\Lambda^{2}}}\right)-\ln^{2}\left(\frac{\beta_{j}}{m_{j}\sqrt{\Lambda^{2}}}\right)+Li_{2}\left(1-\frac{\beta_{i}\left(l_{ij}\cdot\Lambda\right)}{\Lambda^{2}\gamma_{ij}}\right)+Li_{2}\left(1-\frac{m_{i}^{2}\left(l_{ij}\cdot\Lambda\right)}{\beta_{i}\gamma_{ij}}\right)}\nonumber \\
\label{eq:a17}\\
 & {\displaystyle -Li_{2}\left(1-\frac{\beta_{j}\left(l_{ij}\cdot\Lambda\right)}{\alpha_{ij}\Lambda^{2}\gamma_{ij}}\right)-Li_{2}\left(1-\frac{m_{j}^{2}\left(l_{ij}\cdot\Lambda\right)}{\alpha_{ij}\beta_{j}\gamma_{ij}}\right)}\nonumber 
\end{eqnarray}
 where
\begin{eqnarray}
 & {\displaystyle \alpha_{ij}=\frac{\left(k_{i}\cdot k_{j}\right)+\gamma_{ij}}{m_{i}^{2}};} & l_{ij}=\alpha_{ij}k_{i}-k_{j}\label{eq:a18a}\\
 & {\displaystyle \beta_{i}=\left(k_{i}\cdot\Lambda\right)+\sqrt{\left(k_{i}\cdot\Lambda\right)^{2}-m_{i}^{2}\Lambda^{2}};} & {\displaystyle \gamma_{ij}=\sqrt{\left(k_{i}\cdot k_{j}\right)^{2}-m_{i}^{2}m_{j}^{2}}}\label{eq:a18b}
\end{eqnarray}
In the case where $i=j$, the integral in Eq.(\ref{eq:a14}) has the
simple form:
\begin{eqnarray}
 & {\displaystyle I\left(k_{i},k_{i}\right)=\frac{2\pi}{m_{i}^{2}}\left[\ln\left(\frac{4\Delta\epsilon^{2}}{\lambda^{2}}\right)-\frac{2\left(k_{i}\cdot\Lambda\right)}{\sqrt{\left(k_{i}\cdot\Lambda\right)^{2}-m_{i}^{2}\Lambda^{2}}}\cdot\ln\left(\frac{\beta_{i}}{m_{i}\sqrt{\Lambda^{2}}}\right)\right]}\label{eq:a19}\\
\nonumber 
\end{eqnarray}
The momentum $\Lambda$ is determined by using the soft-photon approximation
and is equal to the momentum of undetected particle $\Lambda=k_{5}$.
The $\Delta\varepsilon$ parameter is the upper value of the energy of the emitted
soft photon. If hard-photon bremsstrahlung is considered,
the dependence on $\Delta\varepsilon$ is canceled and replaced by
the experimental cuts. In this work, we have decided to keep only the
soft-photon contribution and determine $\Delta\varepsilon$ from the
pion production threshold conditions. That translates into the requirement
that the invariant mass of the undetected hadron and the emitted soft photon
should not exceed the value of $m_{5}+m_{\pi}$. As a result, the
maximum energy of the emitted soft photon in \textbf{R} frame can be calculated
in the following way:
\begin{eqnarray}
\Delta\varepsilon=m_{\pi}\Bigg(1-\frac{x}{2(1+x)}\Bigg), & \,\,\,\,\text{where}\,\, x=\frac{m_{\pi}}{m_{5}}.\label{eq:n1}
\end{eqnarray}
For the case of $\pi^{0}$ electroproduction, we have $m_{5}=m_{\pi}$,
and therefore $\Delta\varepsilon=\frac{3}{4}m_{\pi}$. For $\pi^{+}$
electroproduction, we have $m_{5}=m_{n}$ and with the help
of the approximation $x\ll1$ we can write $\Delta\varepsilon=m_{\pi}$.

The scalar products relevant to the soft-photon integral are written in
terms of the invariants of \citep{QEDRadCor} and are defined as:

\begin{eqnarray}
 & {\displaystyle \left(k_{1}\cdot k_{5}\right)=\frac{u_{1}-V_{1}}{2}}, & {\displaystyle \left(k_{2}\cdot k_{5}\right)=\frac{S-S_{t}-X}{2}+m_{2}^{2}}\nonumber \\
 & {\displaystyle \left(k_{3}\cdot k_{5}\right)=\frac{u_{3}-V_{3}}{2},} & {\displaystyle \left(k_{4}\cdot k_{5}\right)=\frac{V_{1}-V_{3}+S_{t}}{2}-m_{4}^{2}.}\label{eq:ap3}
\end{eqnarray}
Finally, we can express the soft-photon factor in terms of the evaluated
integral $I\left(k_{i},k_{j}\right)$:
\begin{eqnarray}
 & \delta_{\gamma}= & {\displaystyle -\frac{\alpha}{4\pi^{2}}\cdot\left(-S\cdot I\left(k_{1},k_{2}\right)+V_{1}\cdot I\left(k_{1},k_{4}\right)+X\cdot I\left(k_{3},k_{2}\right)-V_{3}\cdot I\left(k_{3},k_{4}\right)\right)}.\label{eq:a20}
\end{eqnarray}
A combination of the box corrections for any of the cases in Eqs.(\ref{eq:a9}),
(\ref{eq:a10b}) and (\ref{eq:a10e}) and the soft-photon bremsstrahlung
factor in Eq.(\ref{eq:a20}) cancels the photon mass analytically
and makes the total correction free of this nonphysical parameter.

\section{Results}

In this section we provide the analytical results for the box correction
by deriving compact equations using the SPT prescription only, and
show the explicit cancellation of the photon mass $\lambda$. For the
cases of SPMT and FM, analytical expressions for the four-point tensor
integrals are quite lengthy, so for these cases we will only show numerical results. 

Let us define the total SPT differential cross section and the box
correction treated with the bremsstrahlung contribution as:
\begin{eqnarray}
 & d\sigma_{tot}^{SPT}=d\sigma^{SPT}+d\sigma'_{\gamma}=d\sigma_{0}\left(1+\delta_{box}^{SPT}+\delta_{\gamma}\right)=d\sigma_{0}\left(1+\delta_{tot}^{SPT}\right)\label{eq:a21}\\
\nonumber \\
 & \delta_{tot}^{SPT}=\delta_{box}^{SPT}+\delta_{\gamma},\nonumber 
\end{eqnarray}
and separate the infrared-finite part $\delta_{tot}^{F}$ from the
part regularized by the photon mass $\delta_{tot}^{IR}$ in the following
way:
\begin{eqnarray}
 & \delta_{tot}^{SPT}=\delta_{tot}^{IR}+\delta_{tot}^{F} & .\label{eq:a22a}\\
\nonumber 
\end{eqnarray}
Here, we can write $\delta_{tot}^{IR}=\delta_{box}^{IR}+\delta_{\gamma}^{IR}$,
where the infrared-regularized corrections $\delta_{box}^{IR}$ and
$\delta_{\gamma}^{IR}$ have the following simple structure:
\begin{eqnarray}
 & {\displaystyle {\displaystyle \delta_{box}^{IR}=-\frac{\alpha}{\pi}\ln\frac{m_{2}^{2}}{\lambda^{2}}\left[\ln\frac{S}{X}-\ln\frac{V_{1}}{V_{3}}\right]}}\label{eq:a22}\\
\nonumber \\
 & {\displaystyle \delta_{\gamma}^{IR}=-\frac{\alpha}{\pi}\ln\frac{4\Delta\varepsilon{}^{2}}{\lambda^{2}}\left[-\ln\frac{S}{X}+\ln\frac{V_{1}}{V_{3}}\right]}.\label{eq:a23}
\end{eqnarray}
It is clear now that a combination of Eq.(\ref{eq:a22}) and Eq.(\ref{eq:a23})
results in the finite total correction which can be written in the
following way:
\begin{eqnarray}
 & \delta_{tot}^{IR}=\delta_{box}^{IR}+\delta_{\gamma}^{IR}={\displaystyle -\frac{\alpha}{\pi}\ln\frac{m_{2}^{2}}{4\Delta\varepsilon{}^{2}}\left[\ln\frac{S}{X}-\ln\frac{V_{1}}{V_{3}}\right]}\label{eq:a24}
\end{eqnarray}
The infrared-finite part of the correction $\delta_{tot}^{F}=\delta_{box}^{F}+\delta_{\gamma}^{F}$
can be simplified for the box part as: 
\begin{eqnarray}
{\displaystyle {\displaystyle \delta_{box}^{F}=-\frac{\alpha}{\pi}\left[\frac{1}{2}\ln\frac{S}{X}\cdot\ln\frac{S\cdot X}{m_{2}^{4}}+\frac{1}{2}\ln\frac{V_{3}}{V_{1}}\cdot\ln\frac{V_{1}\cdot V_{3}}{m_{4}^{4}}-\pi^{2}-Li_{2}\left(\frac{S+m_{2}^{2}}{S}\right)+Li_{2}\left(\frac{X-m_{2}^{2}}{X}\right)+\right.}}\nonumber \\
\label{eq:a25}\\
{\displaystyle \left.Li_{2}\left(\frac{V_{1}-m_{4}^{2}}{V_{1}}\right)-Li_{2}\left(\frac{V_{3}+m_{4}^{2}}{V_{3}}\right)\right]}\nonumber \\
\nonumber 
\end{eqnarray}
and the soft-photon finite bremsstrahlung part has the following
structure:
\begin{eqnarray}
{\displaystyle {\displaystyle \delta_{\gamma}^{F}=-\frac{\alpha}{\pi}\left[Li_{2}\left(1-\frac{\beta_{2}\cdot\left(u_{1}-V_{1}\right)}{S\cdot m_{5}^{2}}\right)+Li_{2}\left(1-\frac{m_{2}^{2}\cdot\left(u_{1}-V_{1}\right)}{S\cdot\beta_{2}}\right)-Li_{2}\left(1-\frac{\beta_{4}\cdot\left(u_{1}-V_{1}\right)}{V_{1}\cdot m_{5}^{2}}\right)-\right.}}\nonumber \\
\nonumber \\
Li_{2}\left(1-\frac{m_{4}^{2}\cdot\left(u_{1}-V_{1}\right)}{V_{1}\cdot\beta_{4}}\right)-Li_{2}\left(1-\frac{\beta_{2}\cdot\left(u_{3}-V_{3}\right)}{X\cdot m_{5}^{2}}\right)-Li_{2}\left(1-\frac{m_{2}^{2}\cdot\left(u_{3}-V_{3}\right)}{X\cdot\beta_{2}}\right)+\nonumber \\
\nonumber \\
\left.Li_{2}\left(1-\frac{\beta_{4}\cdot\left(u_{3}-V_{3}\right)}{V_{3}\cdot m_{5}^{2}}\right)+Li_{2}\left(1-\frac{m_{4}^{2}\cdot\left(u_{3}-V_{3}\right)}{V_{3}\cdot\beta_{4}}\right)\right]. &  & {\displaystyle }\label{eq:a26}
\end{eqnarray}
Here, functions $\beta_{2,\,4}$ are
\begin{eqnarray}
\beta_{2}=\frac{S-S_{t}-X}{2}+m_{2}^{2}+\sqrt{\left(\frac{S-S_{t}-X}{2}+m_{2}^{2}\right)^{2}-m_{2}^{2}\cdot m_{5}^{2}}\nonumber \\
\nonumber \\
\beta_{4}=\frac{V_{1}-V_{3}+S_{t}}{2}-m_{4}^{2}+\sqrt{\left(\frac{V_{1}-V_{3}+S_{t}}{2}-m_{4}^{2}\right)^{2}-m_{4}^{2}\cdot m_{5}^{2}} & .\label{eq:a27}
\end{eqnarray}
A combination of equations Eqs.(\ref{eq:a22a}), (\ref{eq:a24}),
(\ref{eq:a25}) and (\ref{eq:a26}) represent a rather compact set
of final expressions for the box correction to the pion electroproduction
evaluated using the SPT prescription of the soft-photon approximation.
At this point we have everything ready to proceed to the numerical
analysis.

\section{Numerical Analysis}

\subsection{The $\pi^{0}$ electroproduction}

\begin{figure*}
\begin{centering}
\includegraphics[scale=0.35]{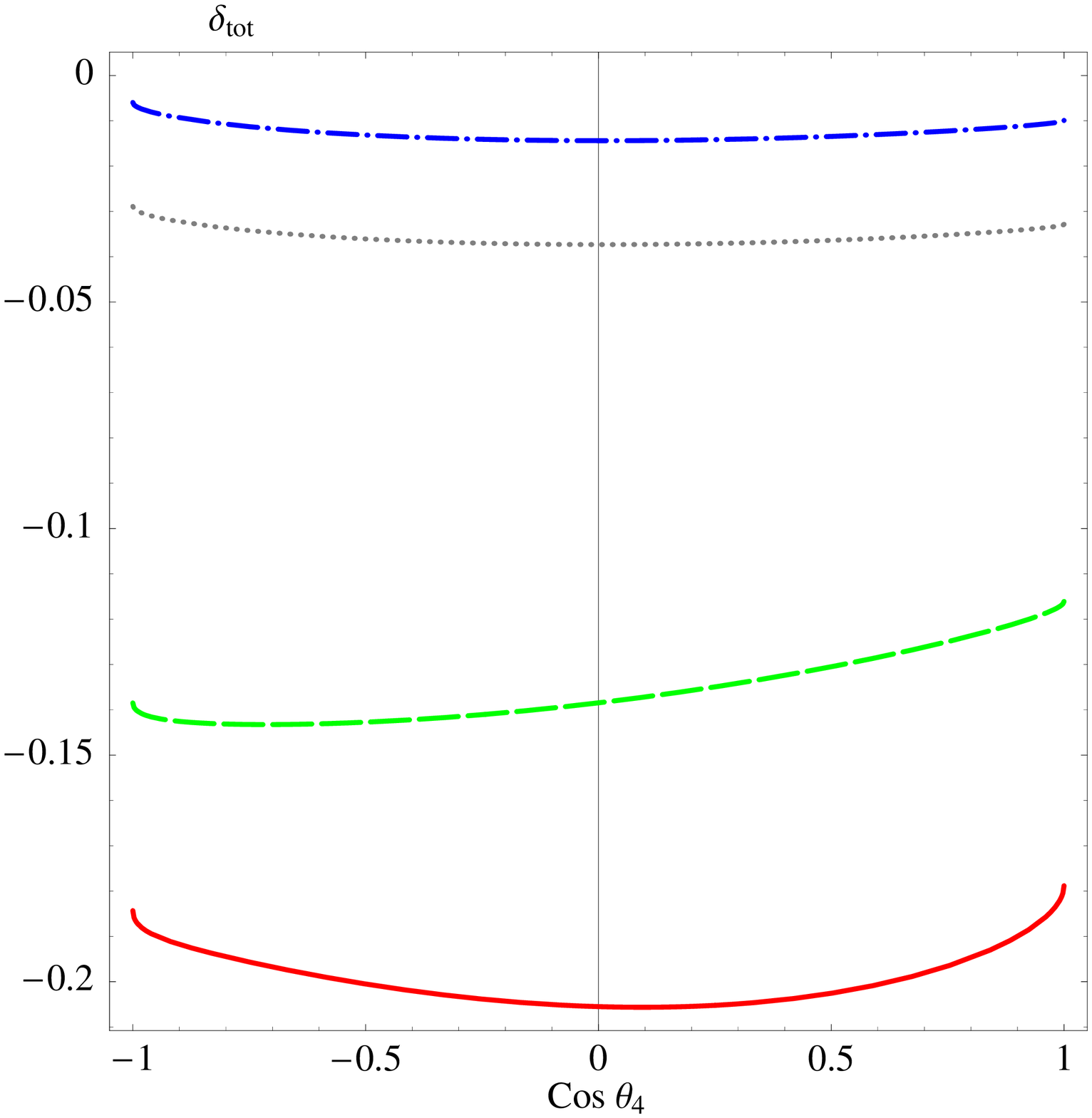}\includegraphics[scale=0.35]{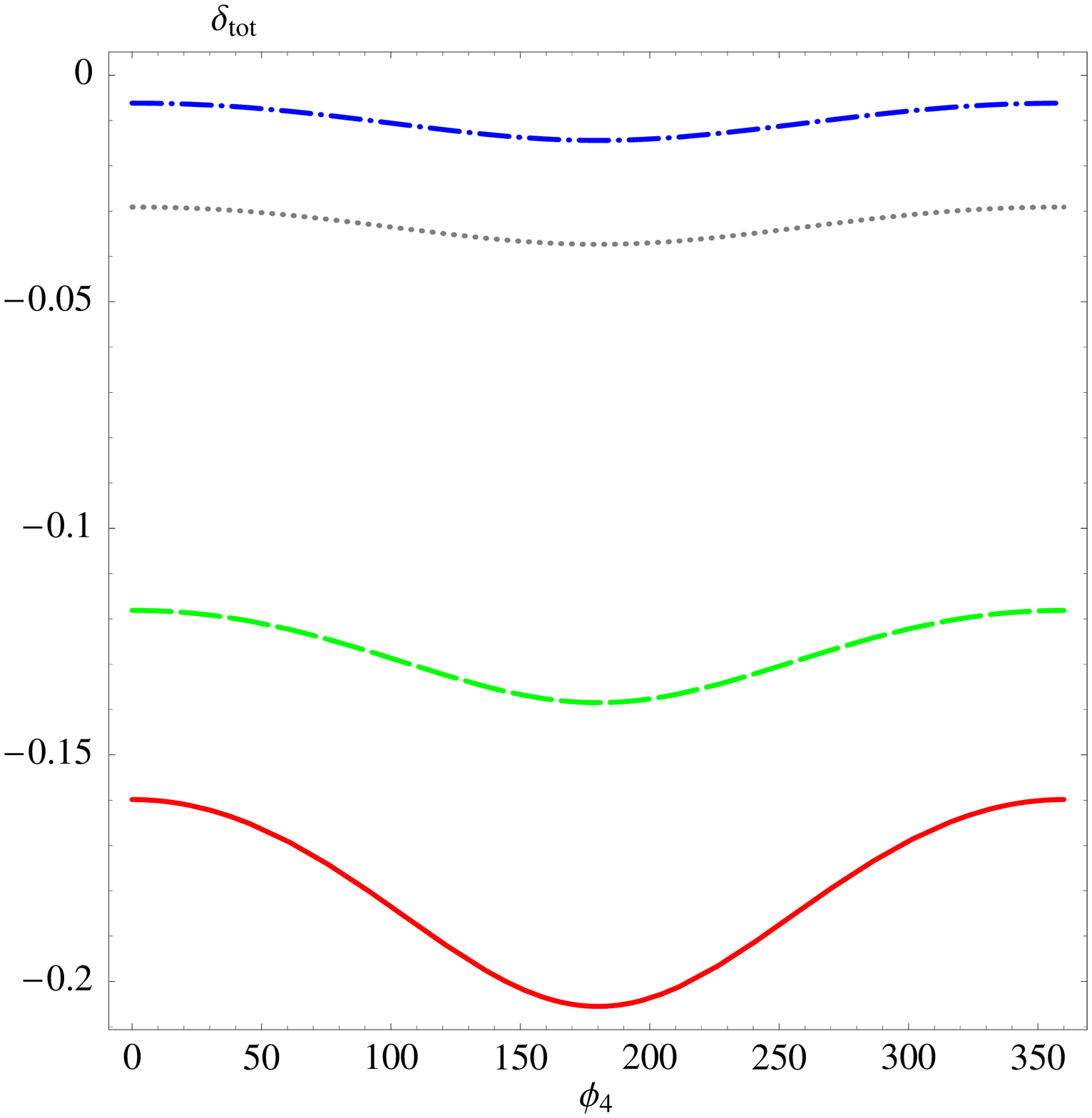}
\par\end{centering}

\begin{centering}
\includegraphics[scale=0.35]{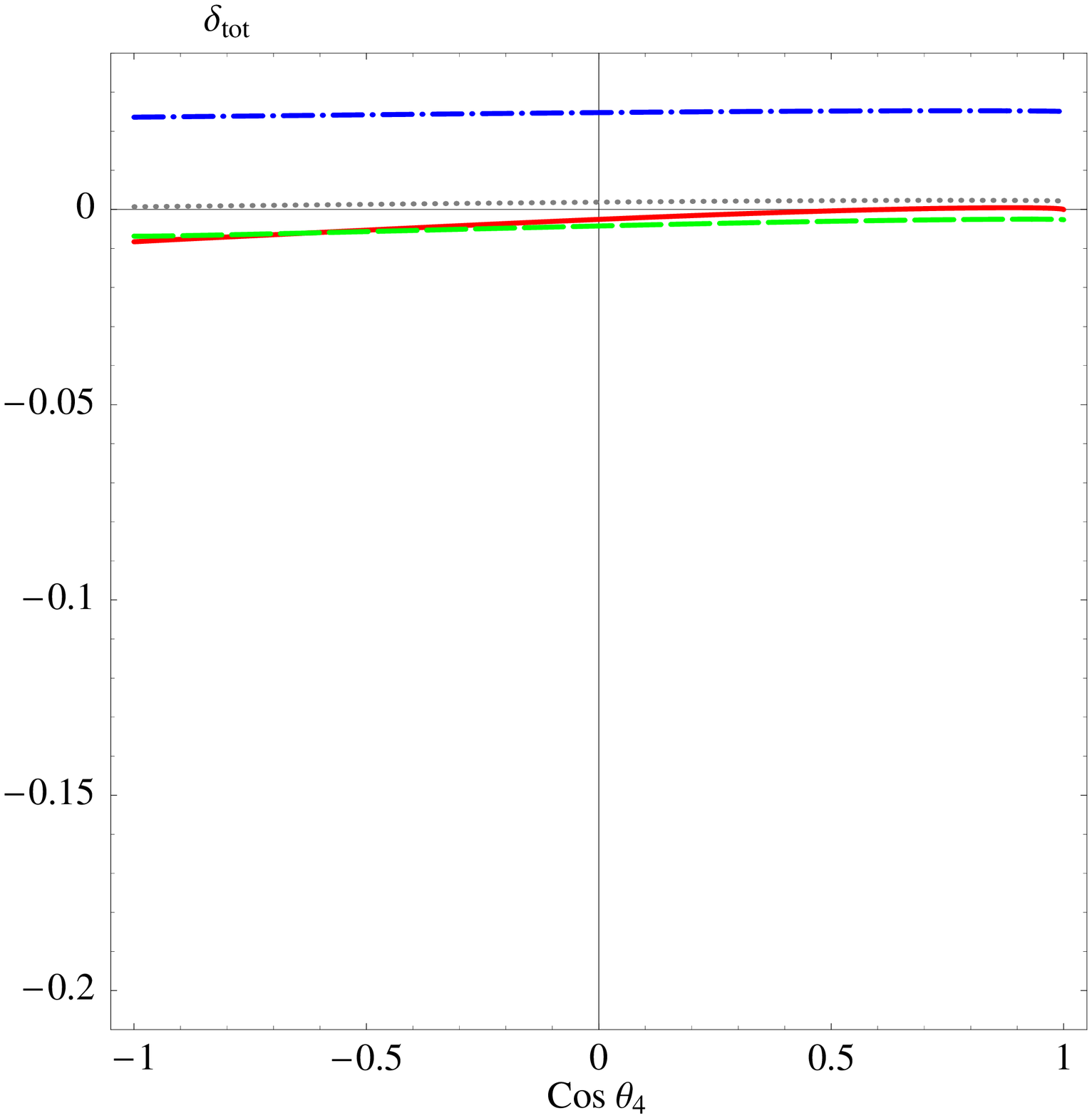}\includegraphics[scale=0.35]{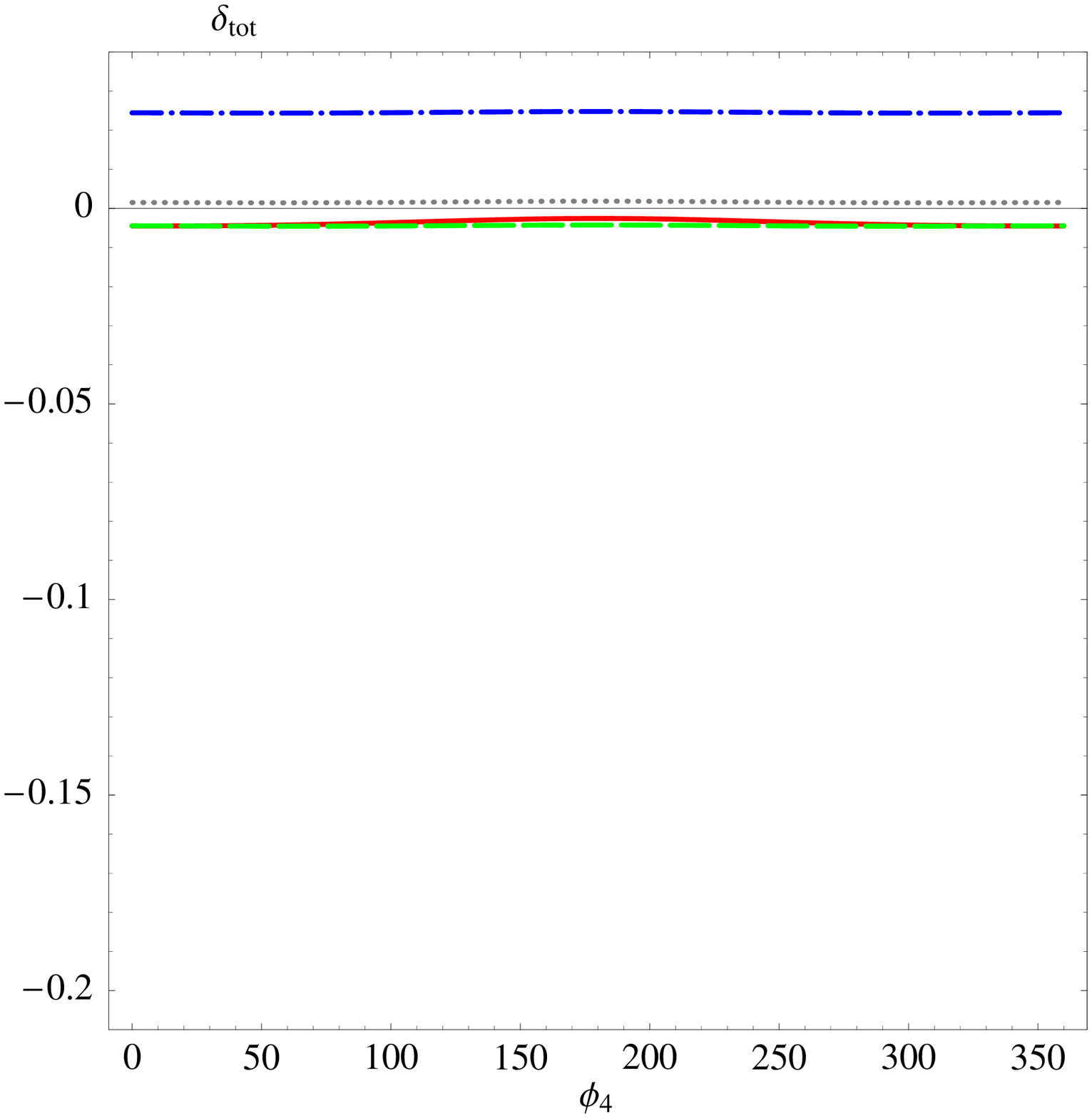}
\par\end{centering}

\caption{$\pi^{0}$ electroproduction two-photon box correction angular
dependencies for the high $Q^{2}=6.36\, GeV^{2}$ (top row) and low
$Q^{2}=0.4\, GeV^{2}$(bottom row) momentum transfers, $W=1.232\, GeV$
and $E_{lab}=5.75\, GeV$. Left column: dependence on $\cos\theta_{4}$
with $\phi_{4}=180^{\circ}$. Right column: dependence on $\phi_{4}$
with $\theta_{4}=90^{\circ}$. Dot-dashed curve - SPT, dotted curve - SPT with $\alpha\pi$ subtracted, dashed curve - SPMT, solid curve - FM approach.}
\label{Flo:pizero1}
\end{figure*}

Let us start with the numerical implementation of the analytical expressions
derived for the two-photon box radiative correction for the kinematics
relevant to the recently completed CLAS \citep{CLAS} experiment performed
in Hall C at JLab. The goal of this experiment \citep{CLAS} is to
measure the $N-\Delta$ transition form factors in the region of high
momentum transfers. Previously, it was found that the radiative correction
arising from the electron vertex corrections, boson self-energies
plus emission of soft and hard photons from an electron could reach as
much as $-23\%$ \citep{Afan2002}. Evidently, these corrections play
a vital role in the extraction of the Born cross section from the
measurements. In addition, the size of the two-photon box radiative
correction could also become an important part of that analysis. Fig.(\ref{Flo:pizero1})
shows our results for the box correction in the region of $\Delta$
resonance as a dependence on the azimuthal and polar angles in the hadronic
center of mass (hadronic kinematic variables) for all approaches considered above and for
the experimental runs with $Q^{2}=0.4,\,6.36\, GeV^{2}$ and electron
beam energy $E_{lab}=5.75\, GeV$ in the lab reference frame. We use the blue dot-dashed line to show the correction
calculated using the SPT prescription. The gray dotted curve represents
the correction obtained in the SPT approach but with subtracted $\alpha\pi$
term. That is equivalent to the result produced by the approximation
$\mbox{Re}\left[C_{0}\left(\{k_{i},m_{i}\},\{-k_{j},m_{j}\}\right)\right]\thickapprox-C_{0}\left(\{k_{i},m_{i}\},\{k_{j},m_{j}\}\right)$,
which was introduced earlier in \citep{Tsai}. The green dashed curve is
a correction obtained using the SPMT method (see \citep{Tjon}). The
red solid line corresponds to the result produced from the exact but
model-dependent FM approach. 

For high momentum transfers the correction in Fig.(\ref{Flo:pizero1})
(top row) shows the strong angular dependencies in the SPMT and FM
approaches and ranges from $-10\%$ to $-20\%$. As for the SPT approach,
the correction is almost constant and lies in the interval of $-0.5\%\thicksim-3.0\%$,
depending on whether we use the approximation of \citep{Tsai} in our calculations.
In general, from Fig.(\ref{Flo:pizero1}) we can see that for high momentum transfer the correction
calculated by the SPT method is notably smaller compared to the SPMT and FM cases. The absolute difference is large and is in the interval of $13\%\thicksim18\%$
for both the polar and azimuthal dependencies. This discrepancy can
be explained by the fact that in the SPMT and FM cases the correction
is calculated using the base of four-point tensor integrals and therefore
it is directly enhanced by the scale of the momentum transfer $\propto-Q^{2}\ln\frac{Q^{2}}{m_{h}^{2}}$
(see Eq.\ref{eq:a10b}). Hence, the difference between the SPT and SPMT, FM approaches is substantially enhanced at higher momentum transfers.
More specifically, for $Q^{2}=0.4\, GeV^{2}$, $E_{lab}=5.75\, GeV$
and $W=1.232\, GeV$, we can see that the difference between all approaches
is smaller and is of the order of $3\%$. This primarily happens due
to the fact that the absolute value of the correction calculated using
the SPMT and FM approaches is substantially decreased for small
momentum transfer (see Fig.(\ref{Flo:pizero1}), bottom row). With
the increase of the invariant mass, we also observe that the correction
tends to become larger at higher momentum transfers.

\begin{figure}
\begin{centering}
\includegraphics[scale=0.34]{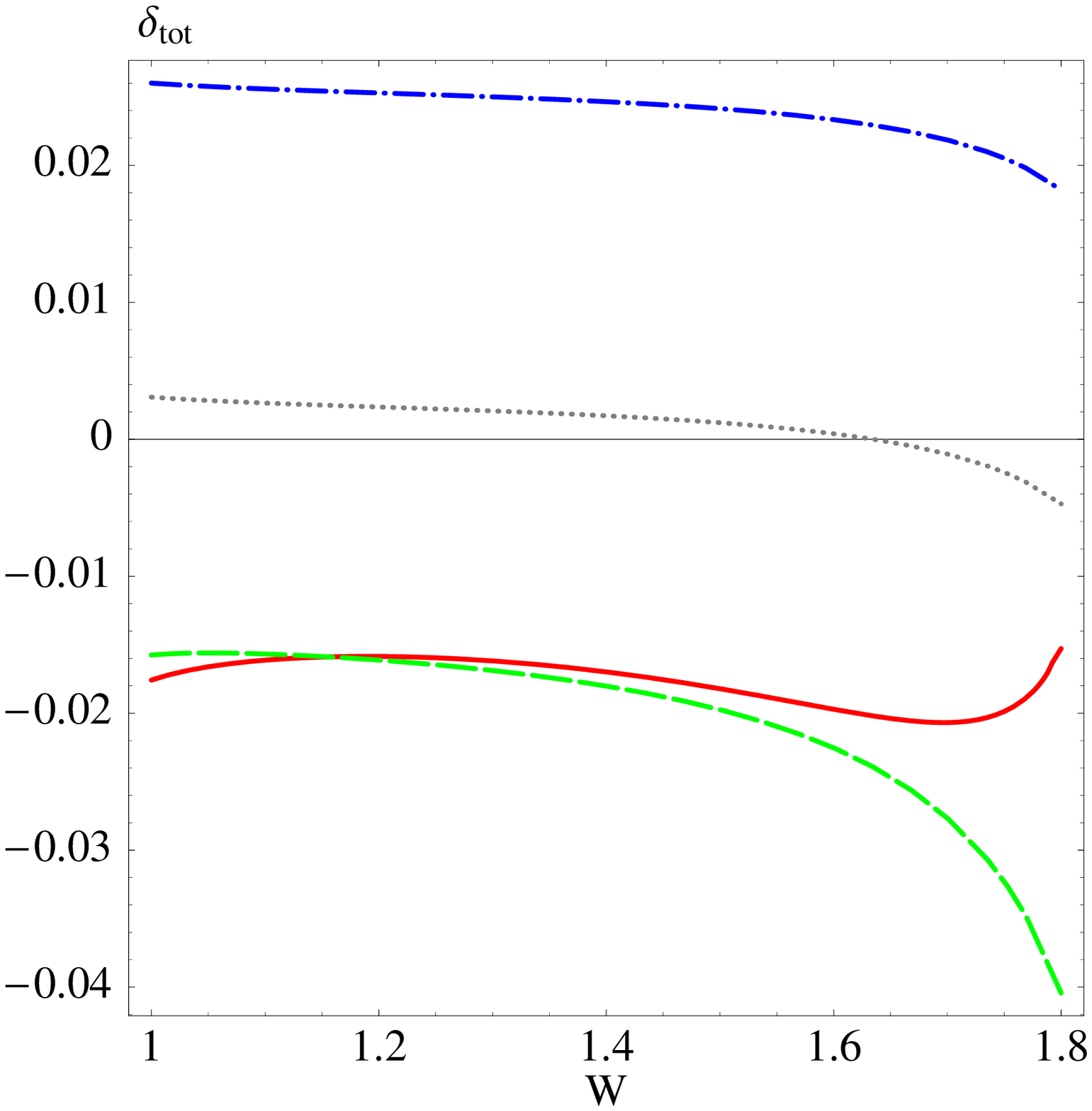}\includegraphics[scale=0.34]{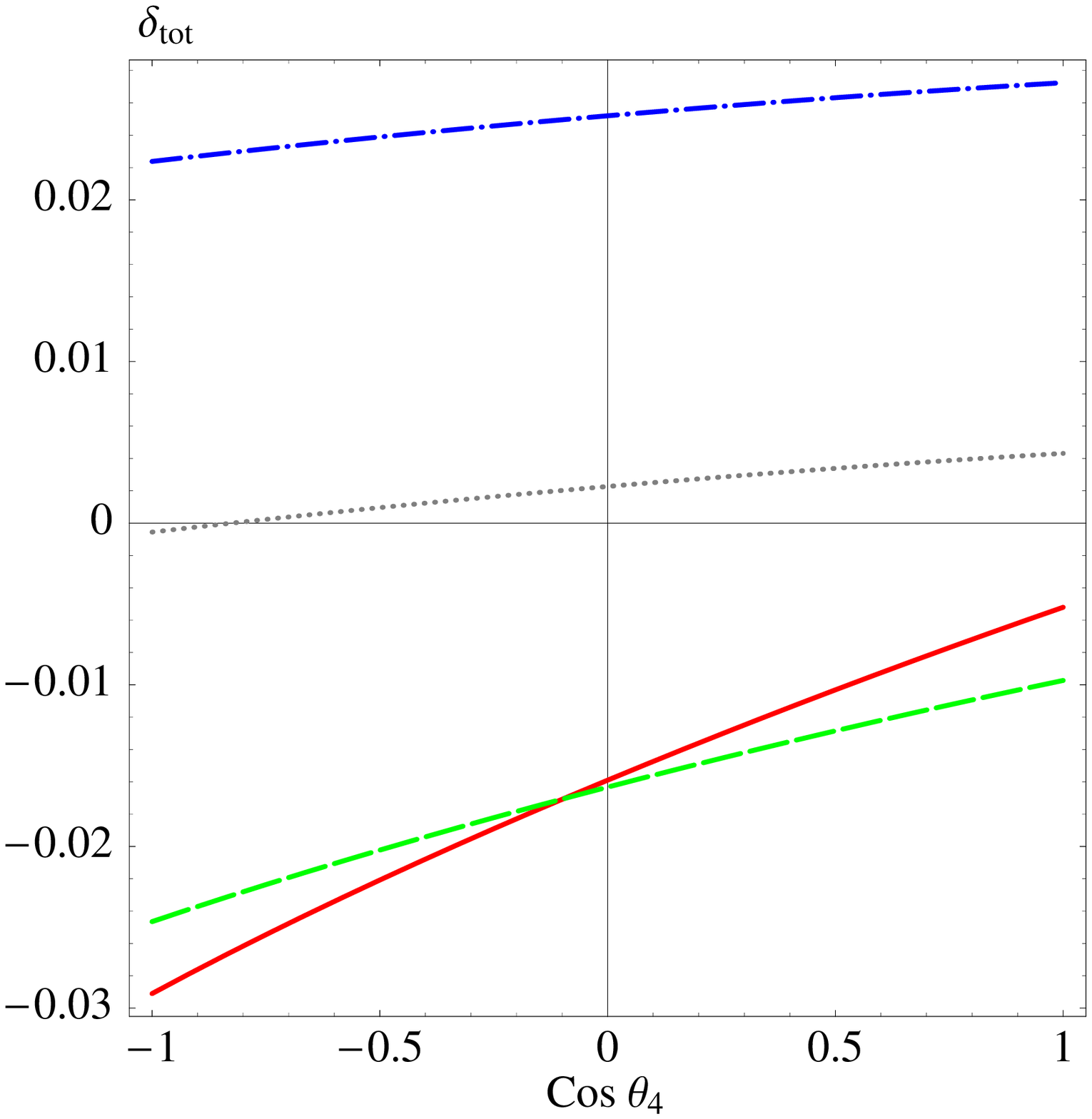}\includegraphics[scale=0.34]{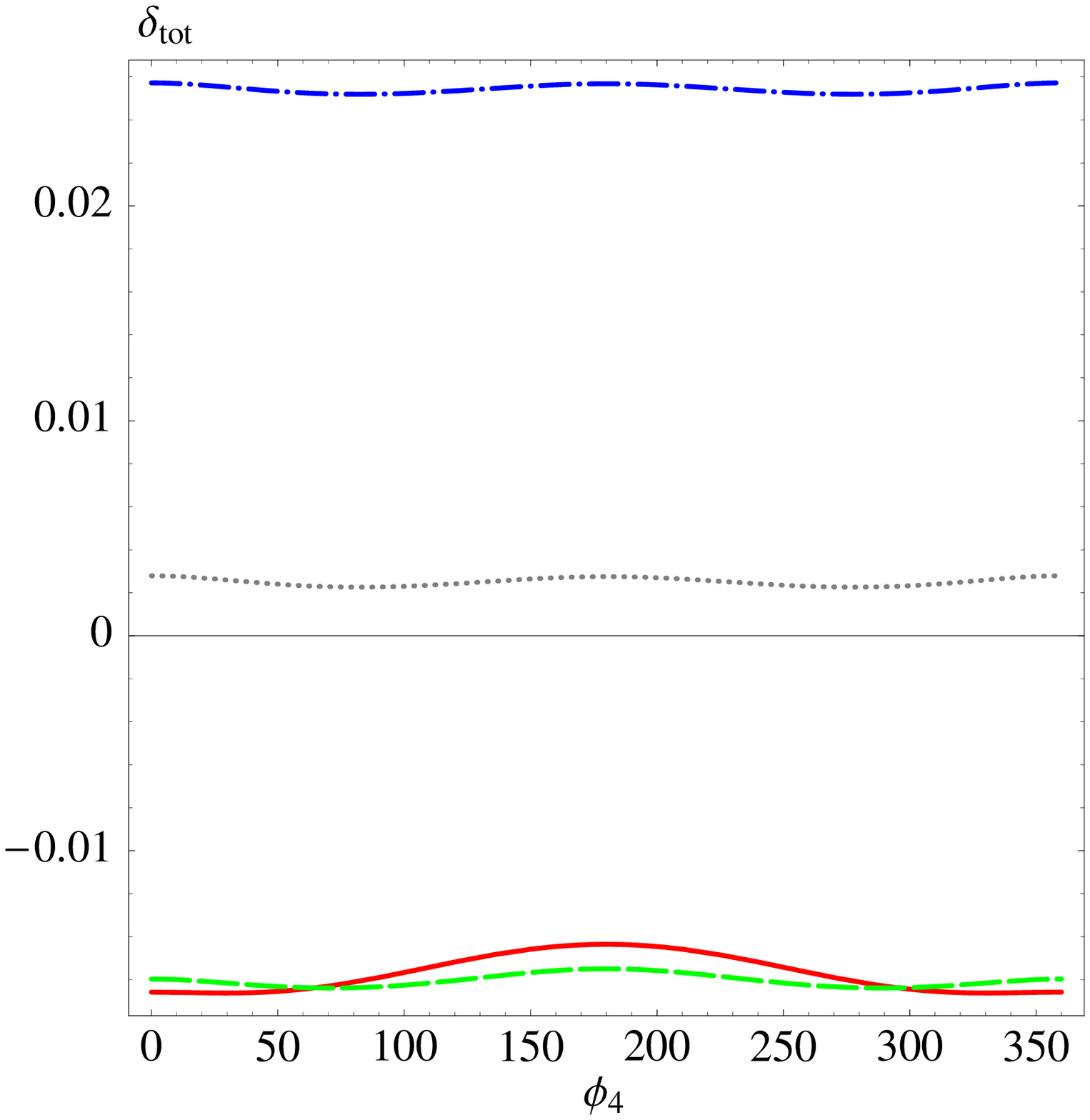}
\par\end{centering}

\caption{$\pi^{0}$ electroproduction box correction for the fixed $Q^{2}=0.4\, GeV^{2}$
and $E_{lab}=1.645\, GeV$. Left plot: dependence on $W$ with $\phi_{4}=90^{\circ}$
and $\theta_{4}=90^{\circ}$. Middle plot: dependence on $\cos\theta_{4}$
with $\phi_{4}=90^{\circ}$ and $W=1.232\, GeV$. Right plot: dependence
on $\phi_{4}$ with $\theta_{4}=90^{\circ}$ and $W=1.232\, GeV$. Dot-dashed curve - SPT, dotted curve - SPT with $\alpha\pi$ subtracted, dashed curve - SPMT, solid curve - FM approach.}
\label{Flo:pizero2}
\end{figure}

Fig.(\ref{Flo:pizero2}), relevant to the earlier CLAS experiments
\citep{CLASa,CLASb,CLASc,CLASd}, shows the low momentum transfer $Q^{2}=0.4\, GeV^{2}$
two-photon box correction dependencies on the invariant mass and the hadronic
polar and azimuthal angles. 
The dependence of the correction on the invariant mass at $Q^{2}=0.4\, GeV^{2}$
is rather weak for all cases with the exception of the SPMT approach.
The angular dependencies show that the correction is rather small and
lies in the interval of $-3.0\%\thicksim2.5\%.$ 

\begin{figure}
\begin{centering}
\includegraphics[scale=0.34]{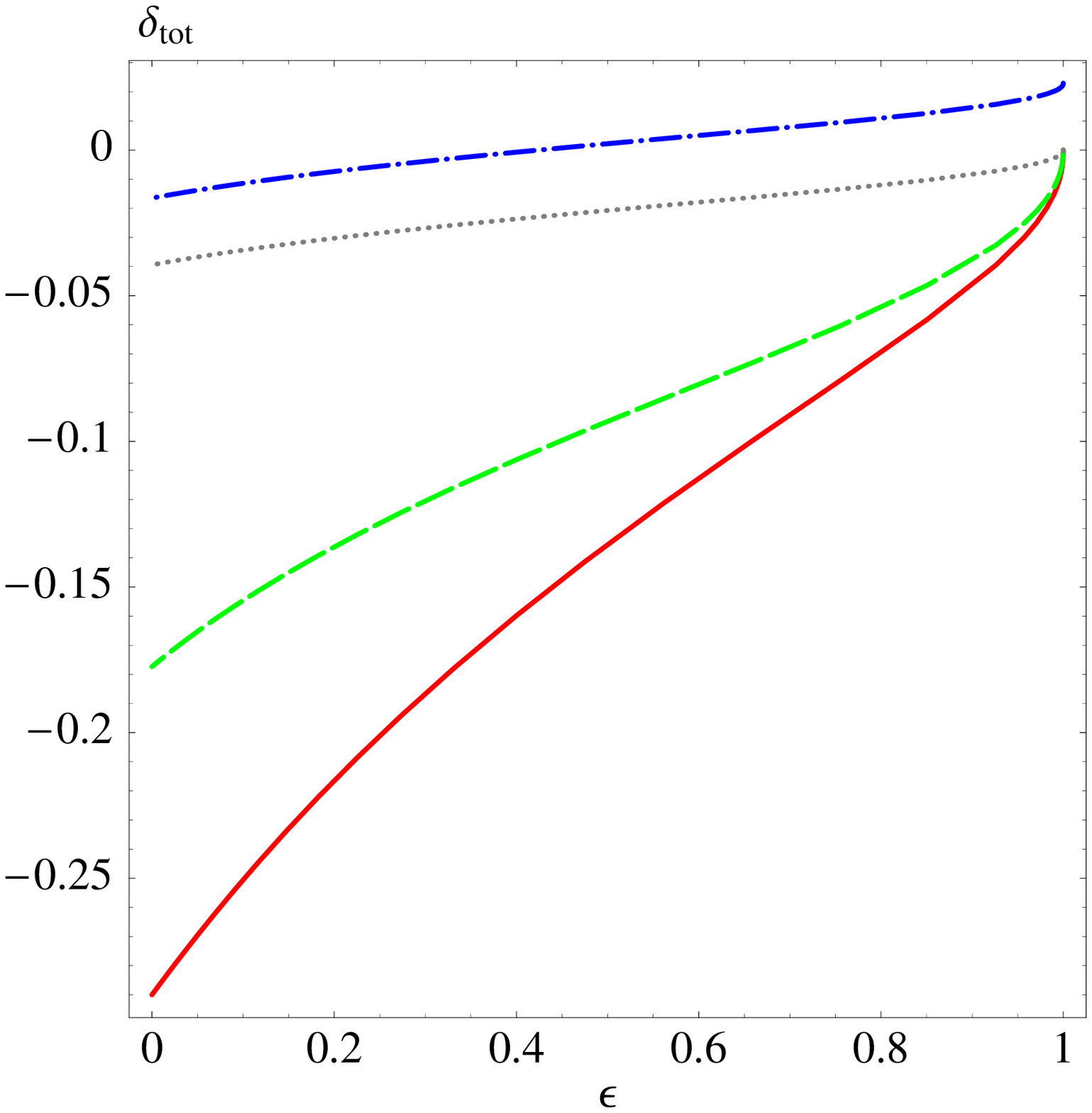}\includegraphics[scale=0.34]{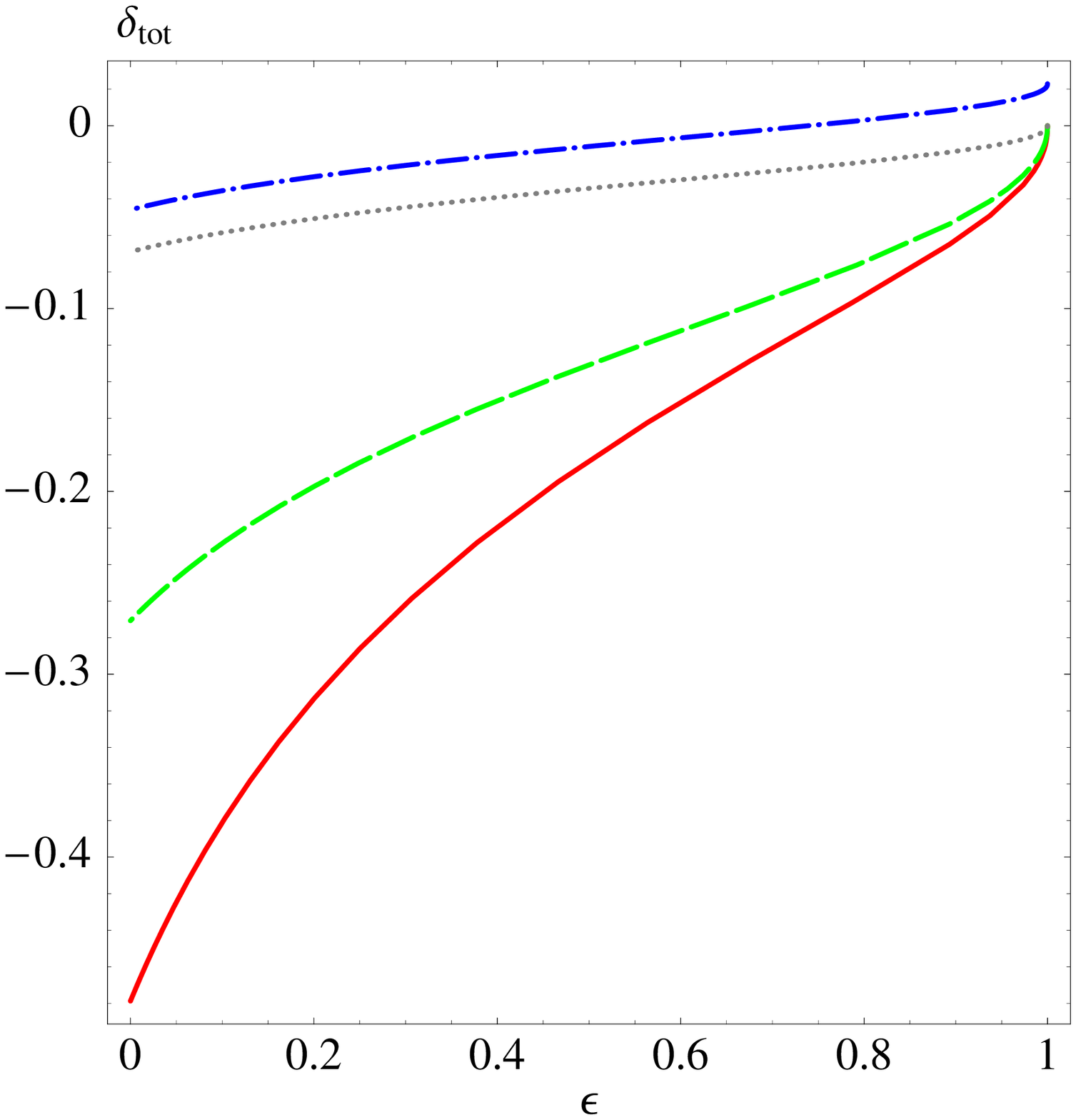}\includegraphics[scale=0.34]{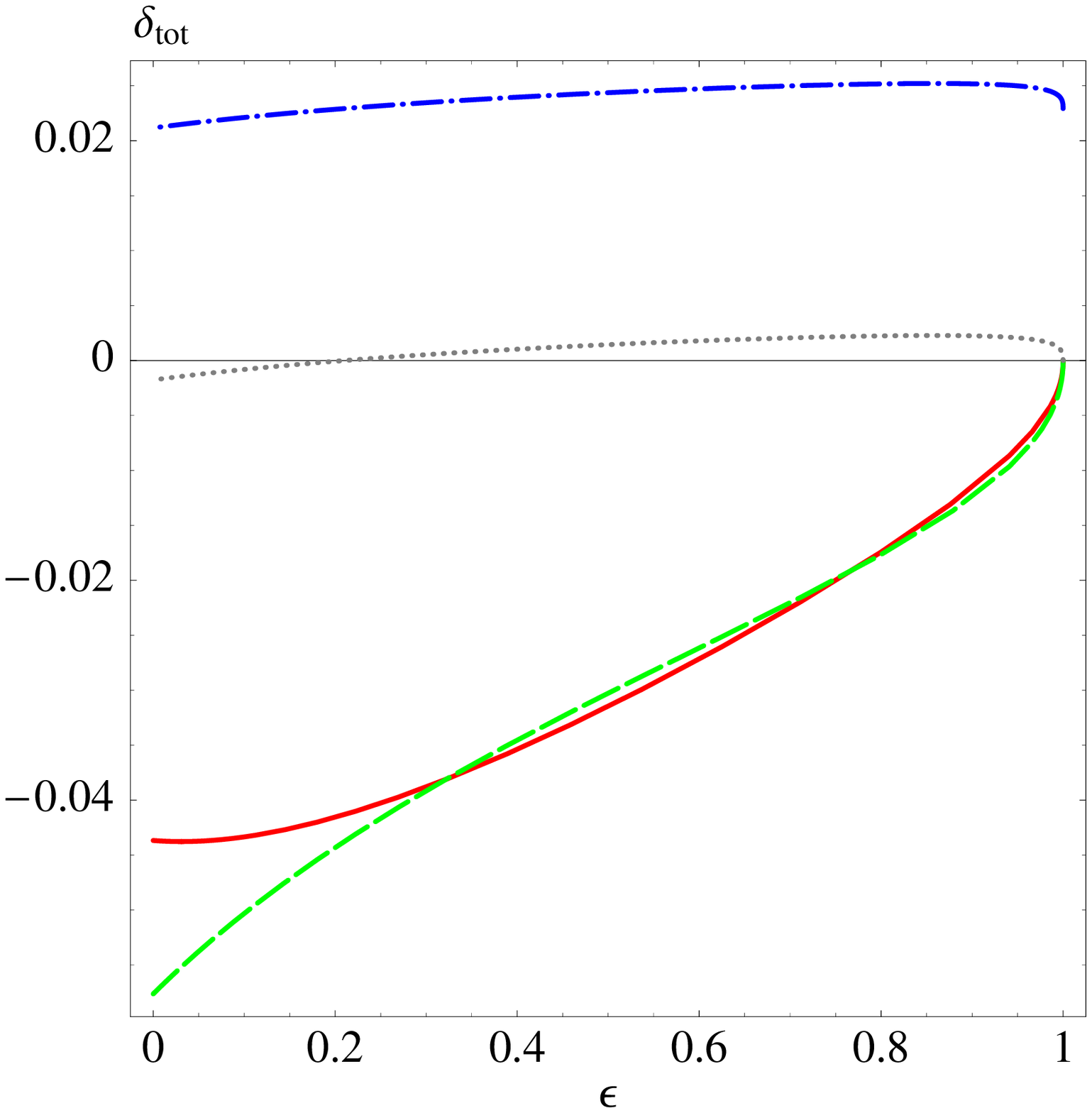}
\par\end{centering}

\caption{$\pi^{0}$ electroproduction two-photon box correction (for detected
proton) dependencies on virtual photon degree of polarization parameter
$\epsilon$ for momentum transfers $Q^{2}=3.0\, GeV^{2}$ (left plot),
$Q^{2}=7.0\, GeV^{2}$ (middle plot) and $Q^{2}=0.4\, GeV^{2}$ (right
plot). All plots are given for $\phi_{4}=90^{\circ}$ and $\theta_{4}=90^{\circ}$
and $W=1.232\, GeV$. Dot-dashed curve - SPT, dotted curve - SPT with $\alpha\pi$ subtracted, dashed curve - SPMT, solid curve - FM approach.}

\label{pizero4}
\end{figure}

\begin{figure}
\begin{centering}
\includegraphics[scale=0.45]{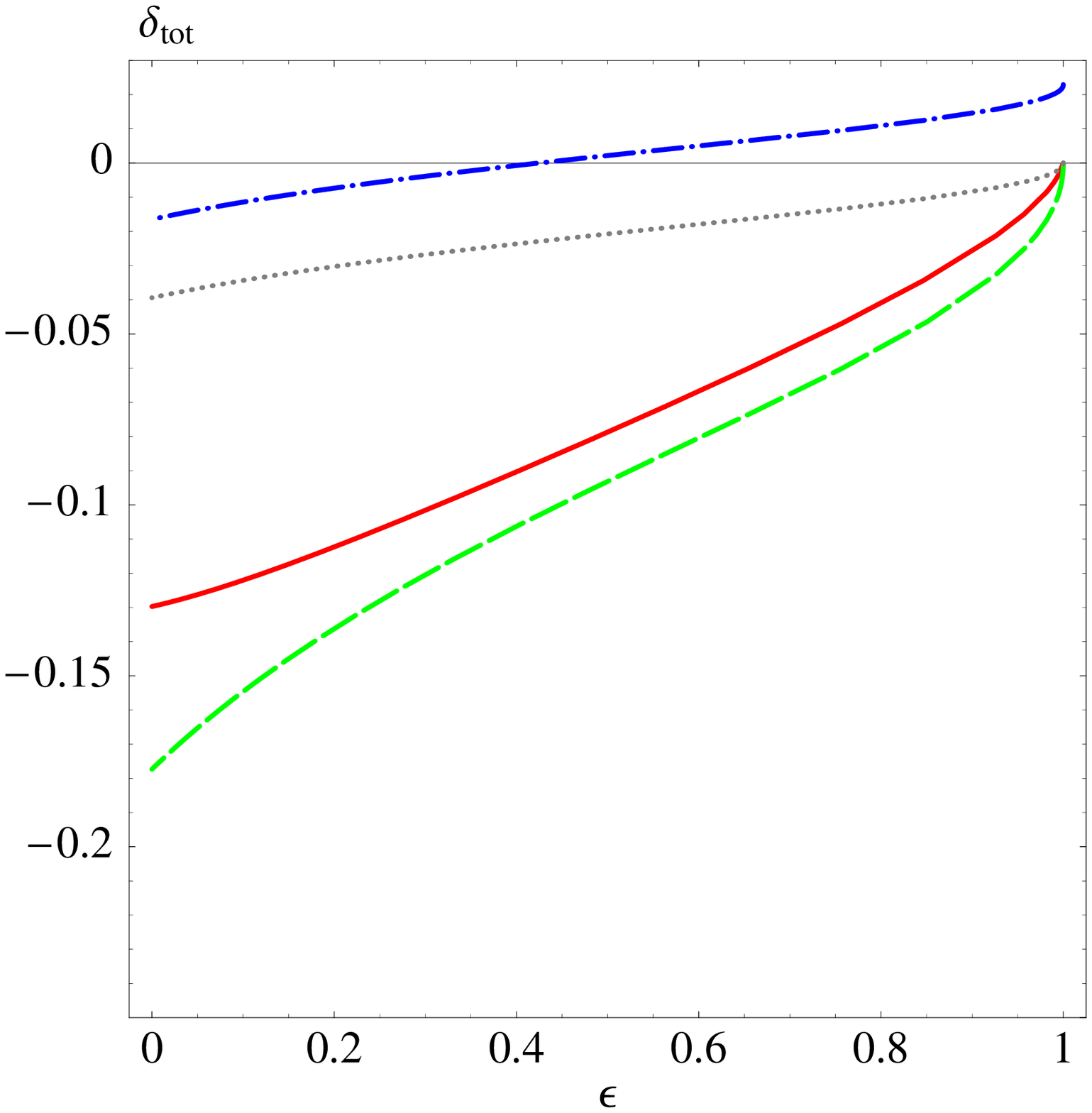}\includegraphics[scale=0.45]{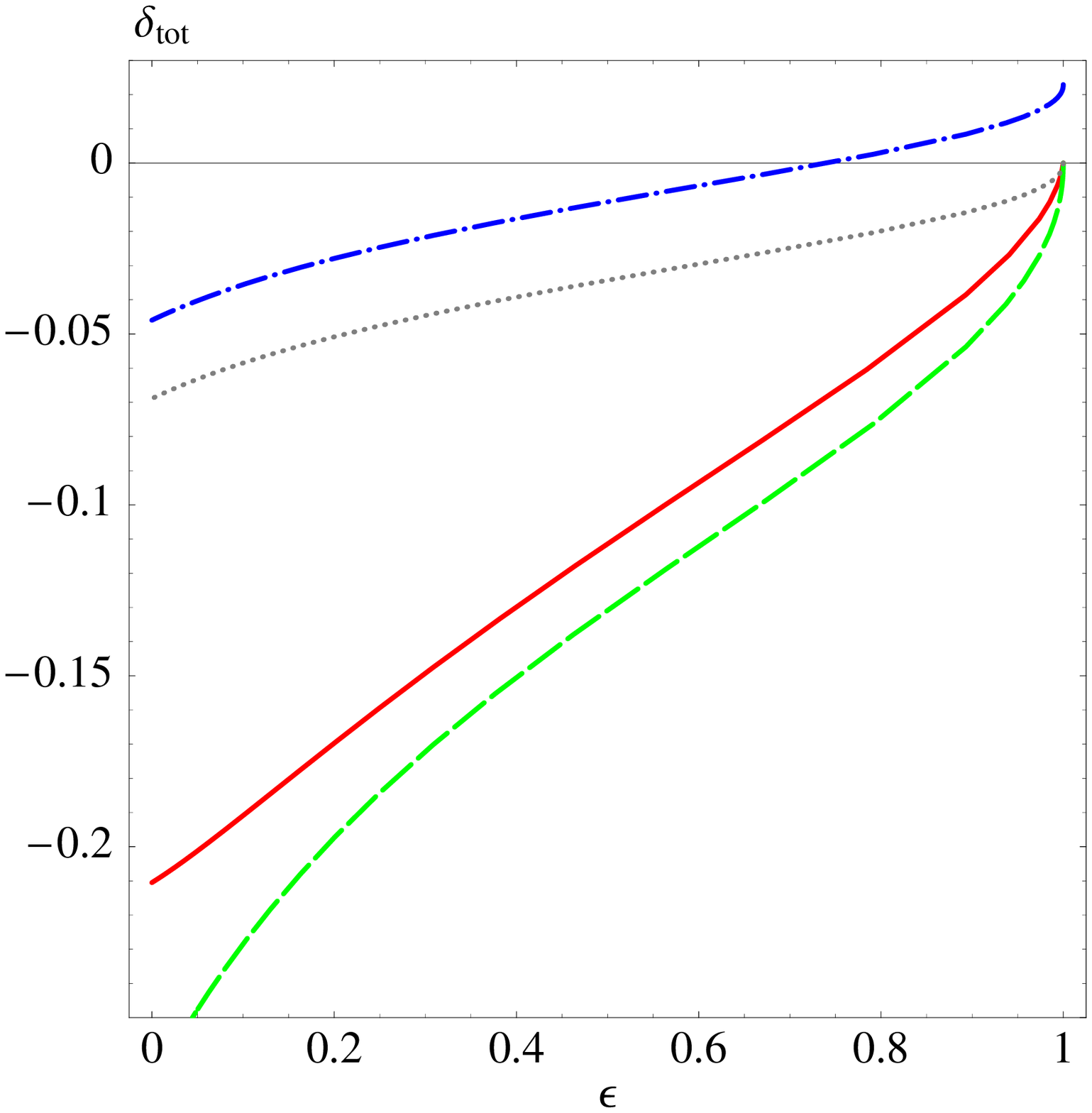}
\par\end{centering}

\caption{$\pi^{0}$electroproduction two-photon box correction dependencies
on the virtual photon degree of polarization $\epsilon$ for the case
of high momentum transfers $Q^{2}=3.0\, GeV^{2}$ (left plot), $Q^{2}=7.0\, GeV^{2}$
(right plot) and without the monopole form factor in the case of exact
calculations (solid line). All plots are given for $\phi_{4}=90^{\circ}$
and $\theta_{4}=90^{\circ}$ and $W=1.232\, GeV$.}

\label{pizero5}
\end{figure}

The study of the radiative correction with respect to the leptonic
kinematic variables can be reproduced by expressing the correction as
a function of the virtual photon degree of polarization parameter
$\epsilon$, which can be evaluated using the following expression:
\begin{eqnarray}
\epsilon=\frac{1}{1+2\ensuremath{\Big(1+\frac{\ensuremath{(S-X)}^{2}}{4Q^{2}m_{2}^{2}}\tan^{2}\theta_{lab}}\Big)}.
\end{eqnarray}
Here, $\theta_{lab}$ is the electron scattering angle in the laboratory
reference frame, which will replace $E_{lab}$ as an input kinematic
parameter. In this case, $E_{lab}$ is no longer an independent
variable and can be calculated in the following way:
\begin{eqnarray}
{\displaystyle E_{lab}=\frac{1}{4m_{2}}\left(Q^{2}+W^{2}-m_{2}^{2}+\sqrt{\big(Q^{2}+W^{2}-m_{2}^{2}\big)^{2}+\frac{4Q^{2}\, m_{2}^{2}}{\sin^{2}\theta_{lab}}}\right)}.
\end{eqnarray}
 For the fixed high momentum transfers of $3.0\, GeV^{2}$ and $7.0\, GeV^{2}$,
the dependence of the two-photon box correction on $\epsilon$ is
shown in Fig.(\ref{pizero4}) (left and middle plots).
Evidently, the box correction calculated in the SPT approach shows
a relatively weak dependence on the degree of polarization parameter
for high momentum transfers and tends to grow for the case of 
backward scattering. As expected, in the SPMT and FM prescriptions,
the correction has a strong dependence on $\epsilon$ and can reach
values of $-17\%\thicksim-50\%$ for backward electron scattering
at $Q^{2}=3.0\thicksim7.0\, GeV^{2}$. It is interesting to observe
that for the high momentum transfer case, the correction calculated in
the region of $\Delta$ resonance using the FM approach has the strongest
dependence on $\epsilon$ and the largest value compared to the results obtained with the SPT
and SPMT prescriptions. This effect could be explained by
the fact that for the high momentum transfers and invariant masses
in the region of $\Delta$ resonance, the two-photon box amplitude evaluated in
the FM approach becomes insensitive to the monopole
form factor. Since we have to divide the interference term in the numerator
of the box correction ($\delta_{box}^{FM}=\frac{2Re[M_{box}^{FM}M_{0}^{\dagger}]}{|M_{0}|^{2}}$)
by $|M_{0}|^{2}$, this effectively produces an overall enhancement
of the correction by a factor of $\frac{\Lambda^{2}+Q^{2}}{\Lambda^{2}}$.
In order to see if that is in fact true, we evaluate the two-photon box correction exactly but without the monopole form factor in
the couplings, hence treating the nucleon and the pion as point-like particles
(see Fig.(\ref{pizero5})). A reduction in the correction
value due to the point-like nucleon and pion would point out that the two-photon
box amplitude calculated in the FM approach is not sufficiently suppressed
by the presence of the monopole fromfactor in order to overcome the $\frac{\Lambda^{2}+Q^{2}}{\Lambda^{2}}$
enhancement coming from the denominator of the box correction.

As can be seen from Fig.(\ref{pizero5}), the correction calculated using the exact approach with the point-like nucleon
and pion is reduced substantially (it is now even smaller than the correction obtained with the SPMT approach), hence justifying the behavior of the correction
calculated in the FM approach at the high momentum transfers. For the
low momentum transfer $Q^{2}=0.4\, GeV^{2}$ (see Fig.(\ref{pizero4})
(right plot)), the correction becomes smaller and all three approaches
produce similar results in the range of $-5.0\%\thicksim2.0\%$.

\subsection{$\pi^{+}$ electroproduction}

\begin{figure}
\begin{centering}
\includegraphics[scale=0.34]{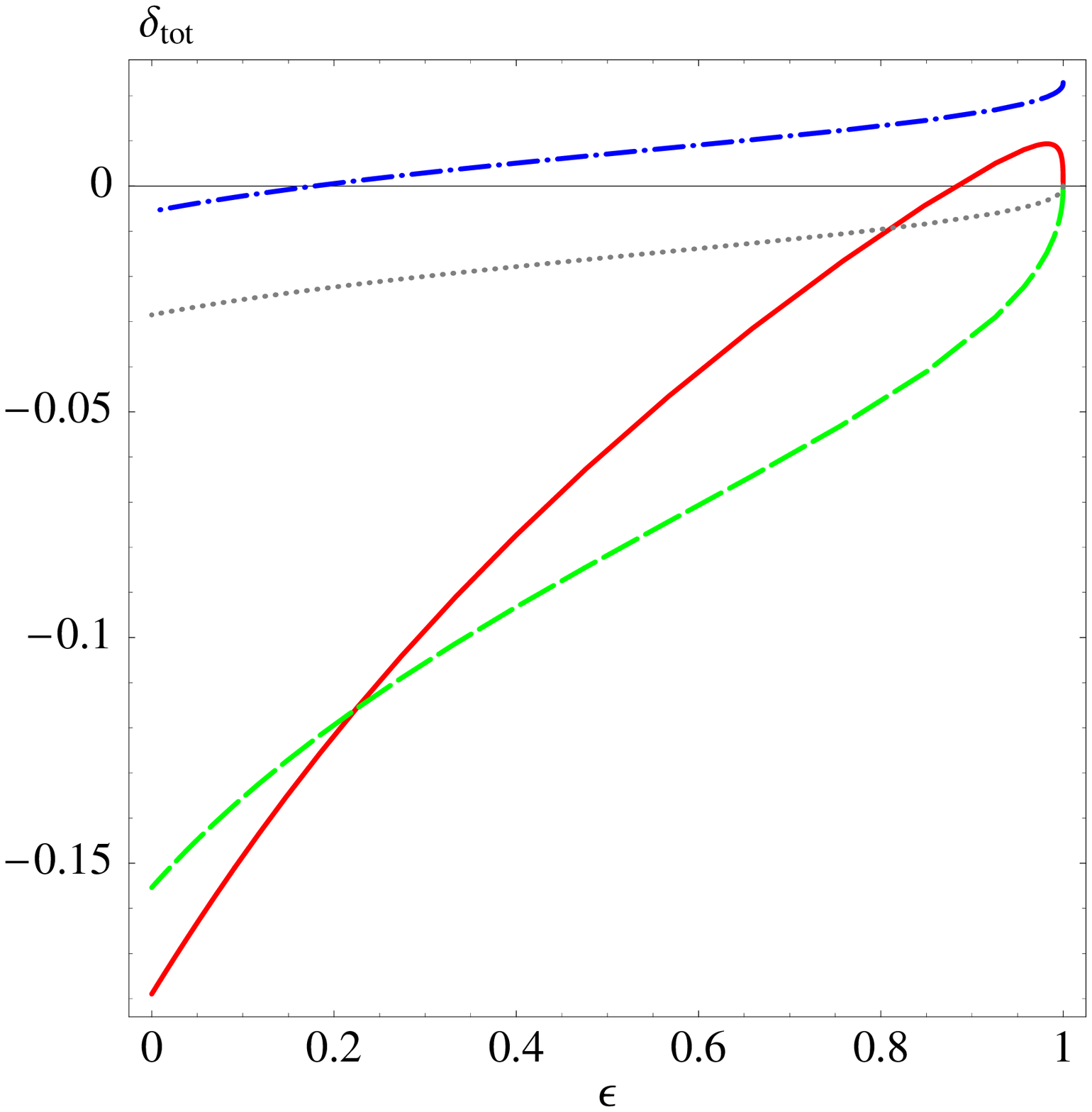}\includegraphics[scale=0.34]{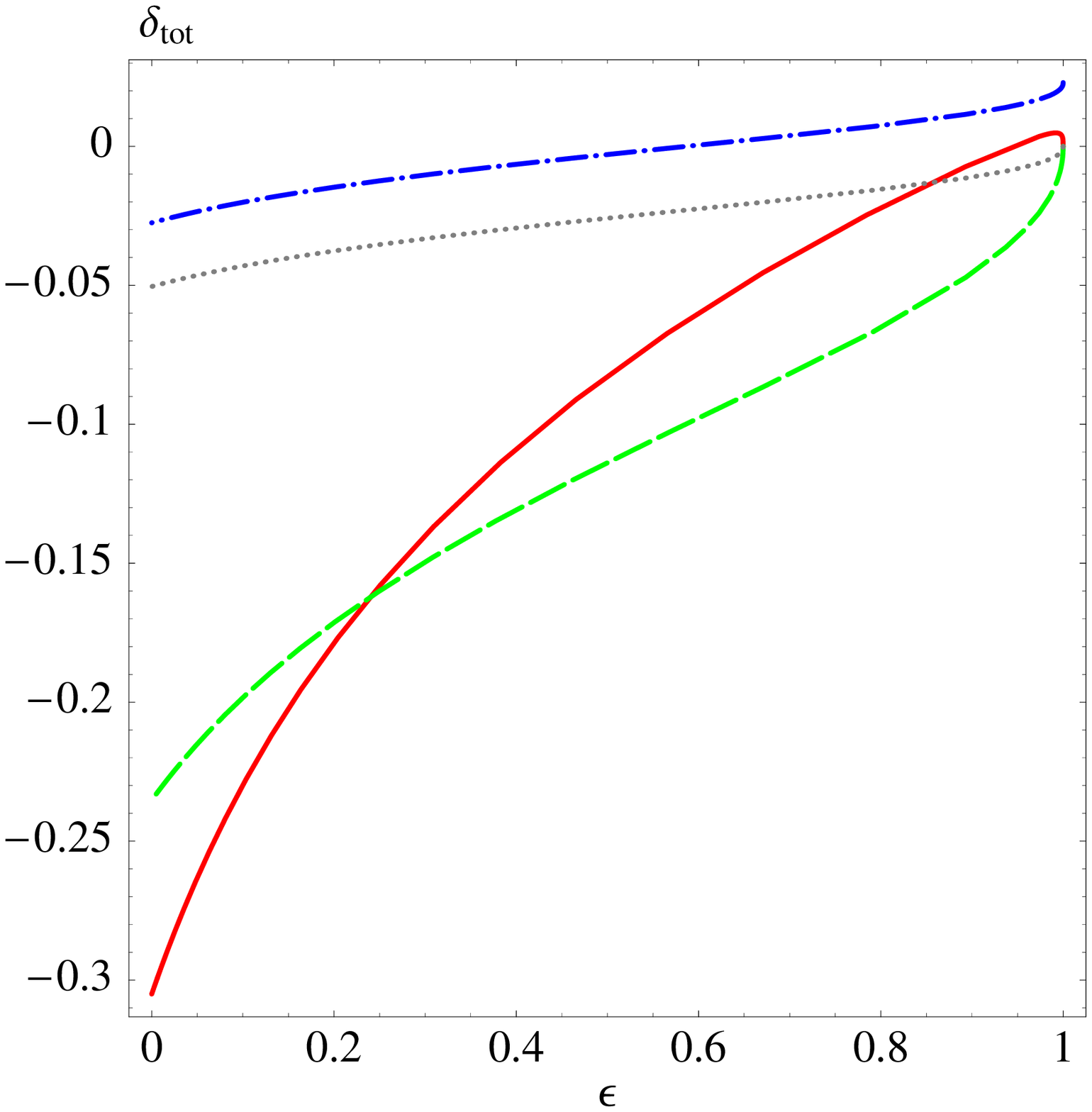}\includegraphics[scale=0.34]{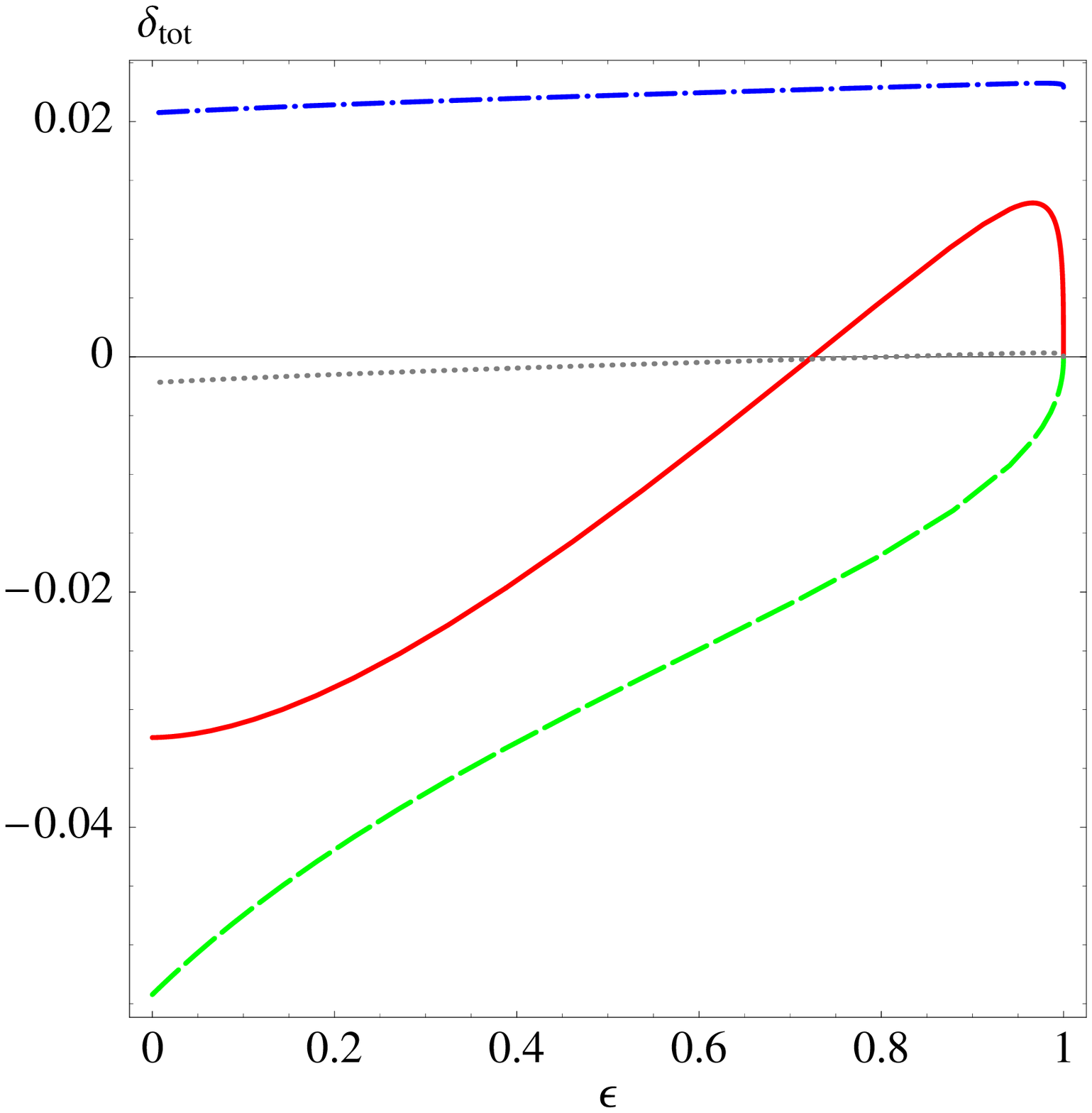}
\par\end{centering}

\caption{Two-photon box correction (for detected $\pi^{+}$) dependencies on
virtual photon degree of polarization parameter $\epsilon$ for $Q^{2}=3.0\, GeV^{2}$
(left plot), $Q^{2}=7.0\, GeV^{2}$ (middle plot) and $Q^{2}=0.4\, GeV^{2}$
(right plot). All plots are given for $\phi_{4}=90^{\circ}$ and $\theta_{4}=90^{\circ}$
and $W=1.232\, GeV$. Dot-dashed curve - SPT, dotted curve - SPT with $\alpha\pi$ subtracted, dashed curve - SPMT, solid curve - FM approach.}

\label{piplus1a}
\end{figure}

\begin{figure}
\begin{centering}
\includegraphics[scale=0.34]{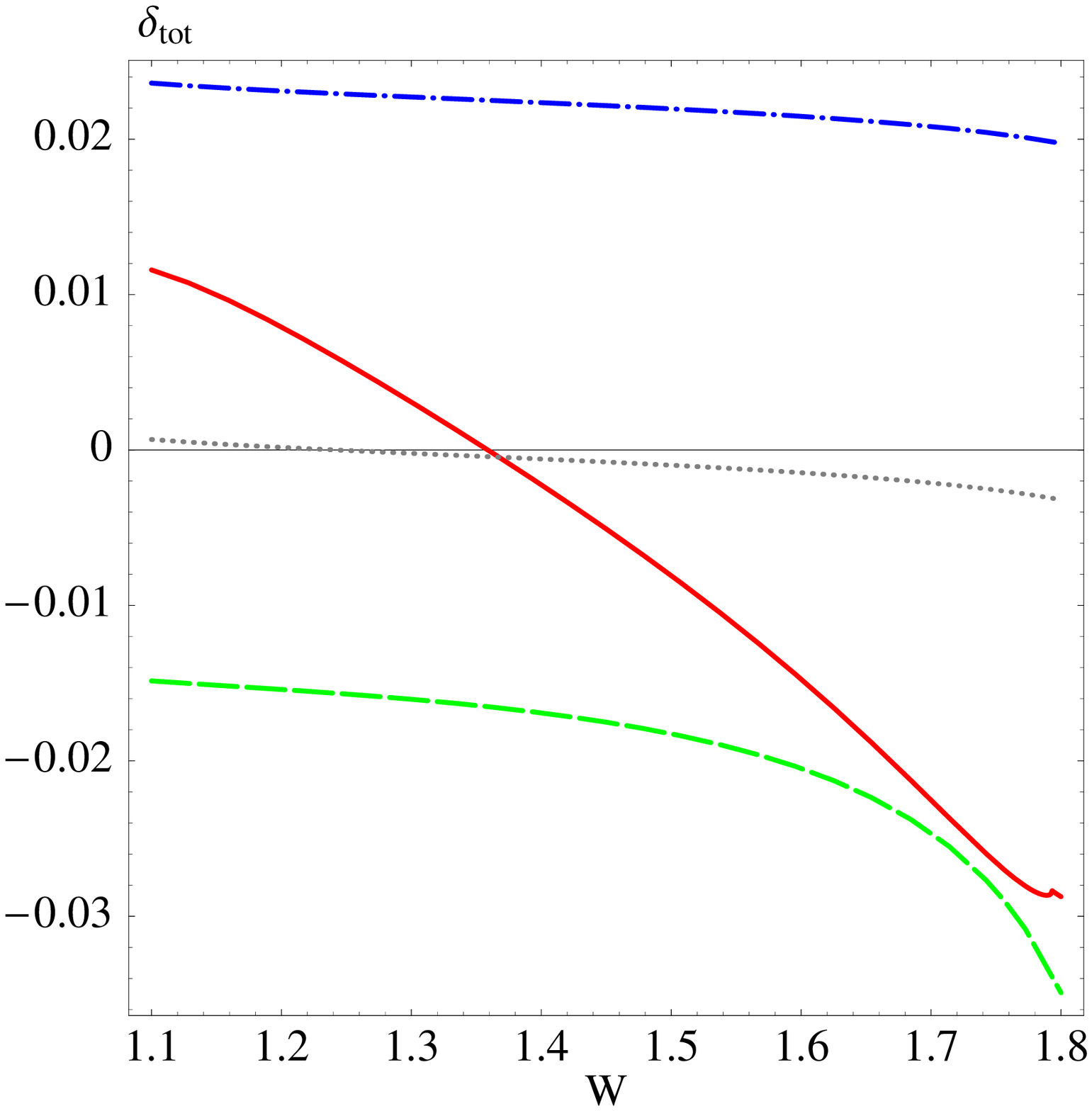}\includegraphics[scale=0.34]{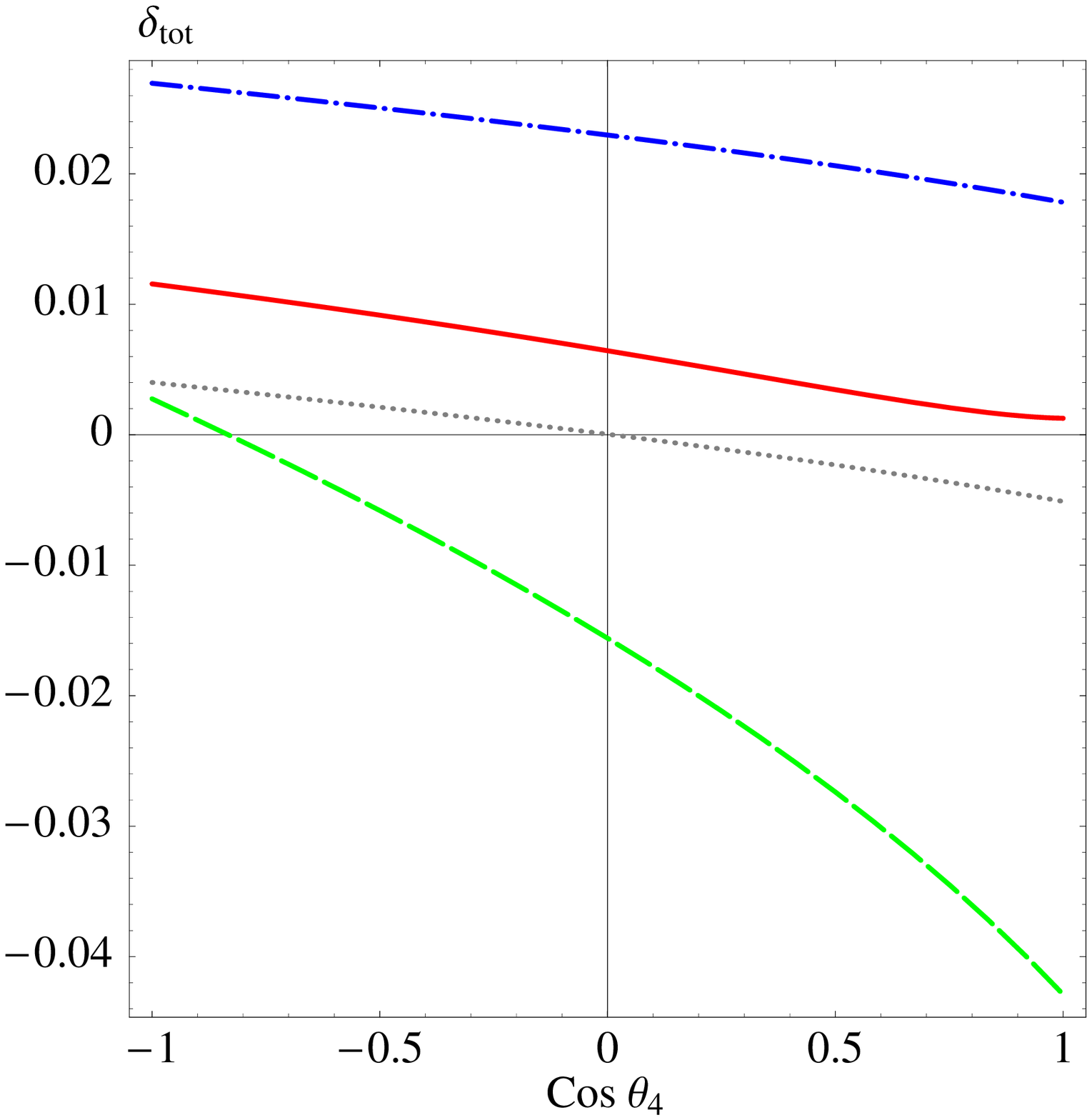}\includegraphics[scale=0.34]{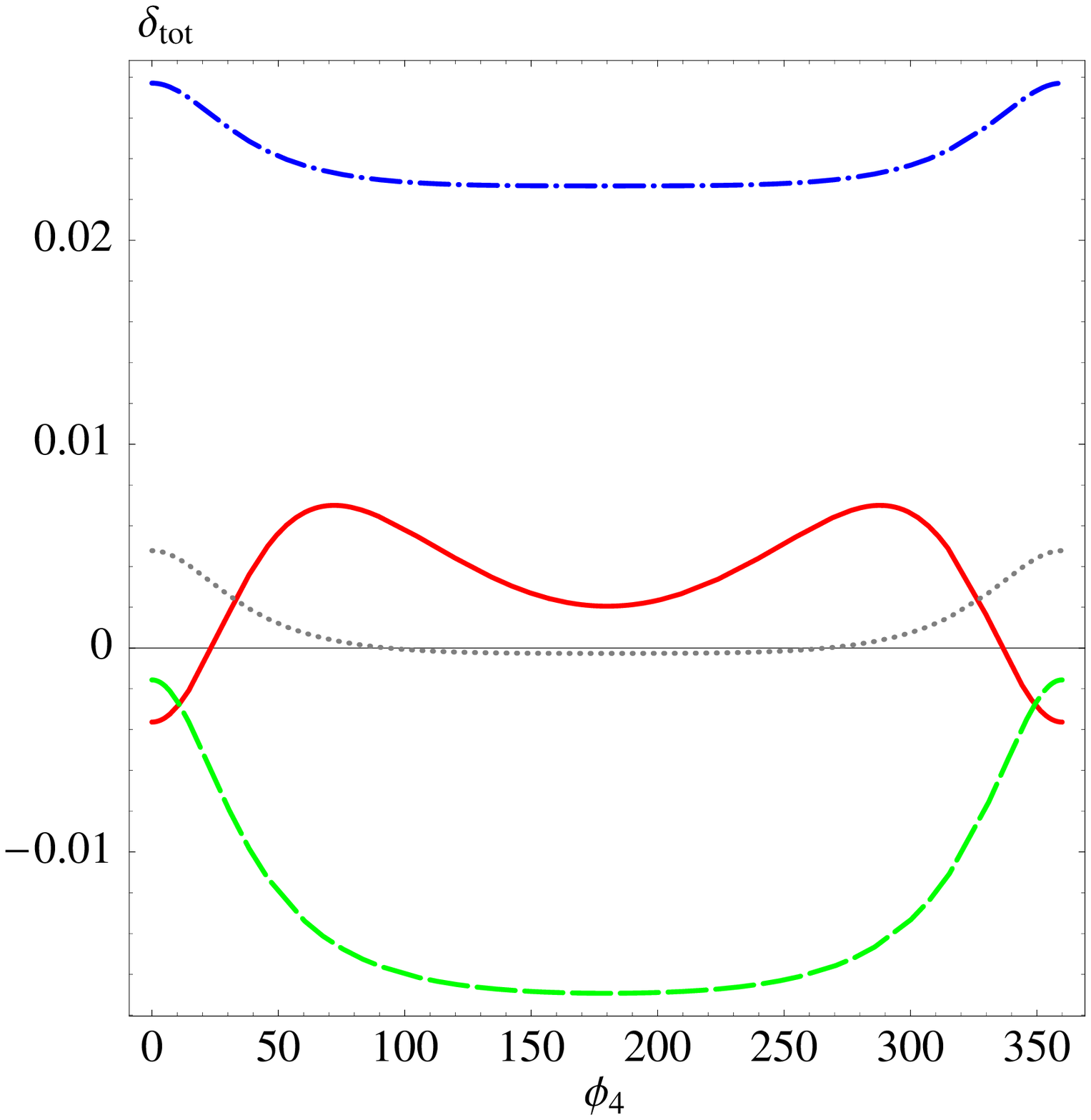}
\par\end{centering}

\caption{$\pi^{+}$ electroproduction box correction for fixed $Q^{2}=0.4\, GeV^{2}$
and $E_{lab}=1.645\, GeV$. Left plot: dependence on $W$ with $\phi_{4}=90^{\circ}$
and $\theta_{4}=90^{\circ}$. Middle plot: dependence on $\cos\theta_{4}$
with $\phi_{4}=90^{\circ}$ and $W=1.232\, GeV$. Right plot: dependence
on $\phi_{4}$ with $\theta=90^{\circ}$ and $W=1.232\, GeV$. Dot-dashed curve - SPT, dotted curve - SPT with $\alpha\pi$ subtracted, dashed curve - SPMT, solid curve - FM approach.}

\label{Flo:piplus2}
\end{figure}

\begin{figure}
\begin{centering}
\includegraphics[scale=0.34]{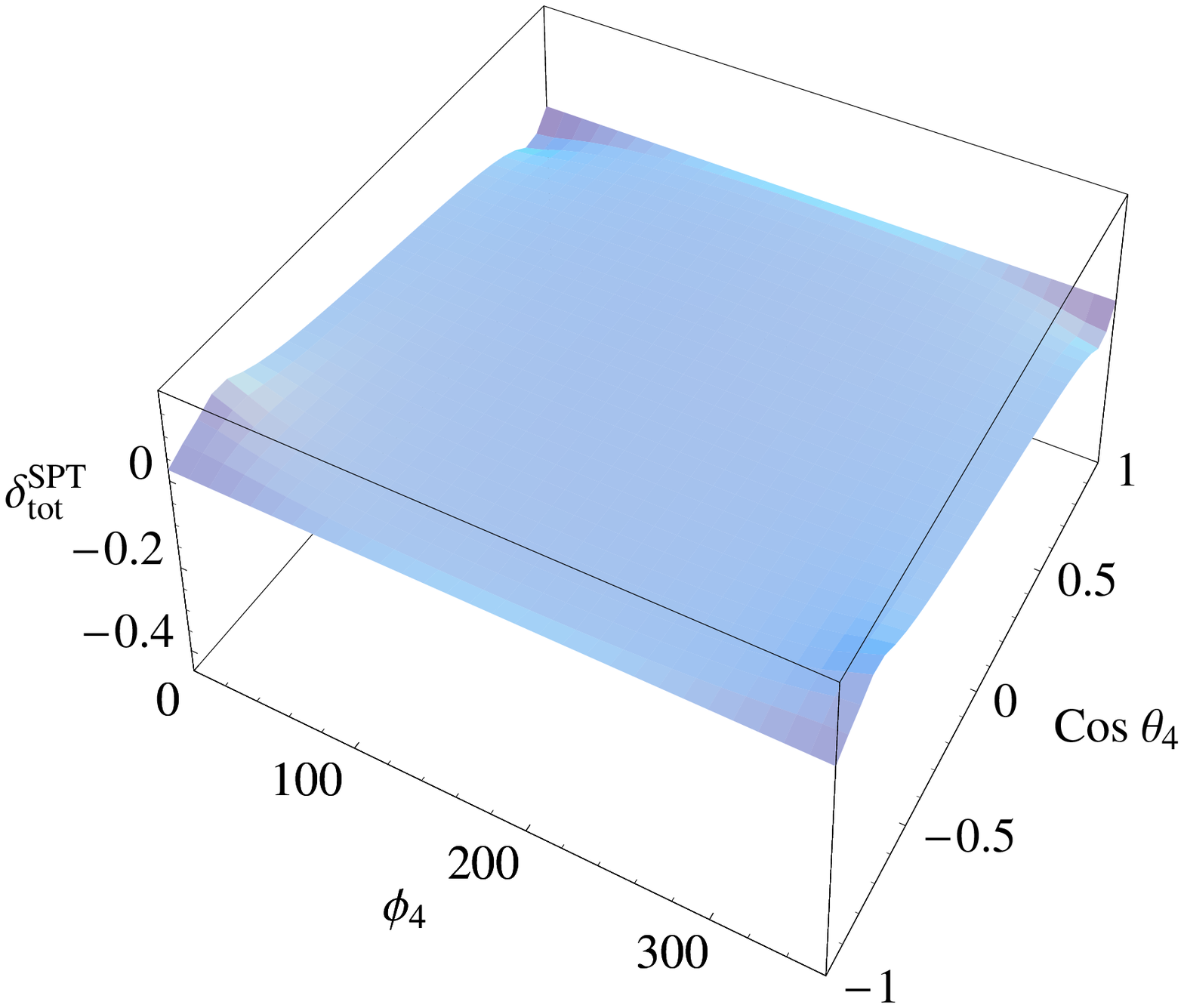}~\includegraphics[scale=0.34]{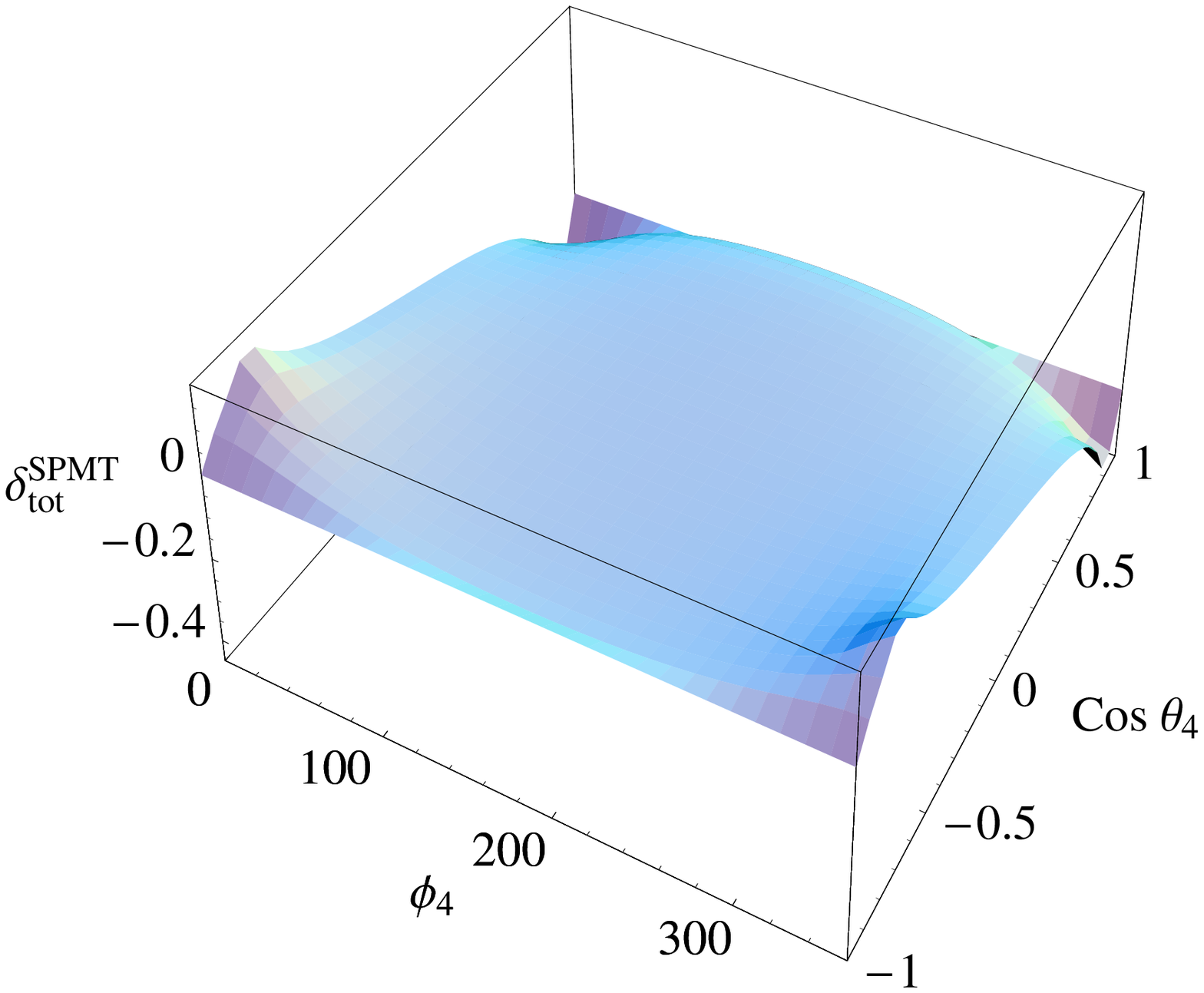}\includegraphics[scale=0.34]{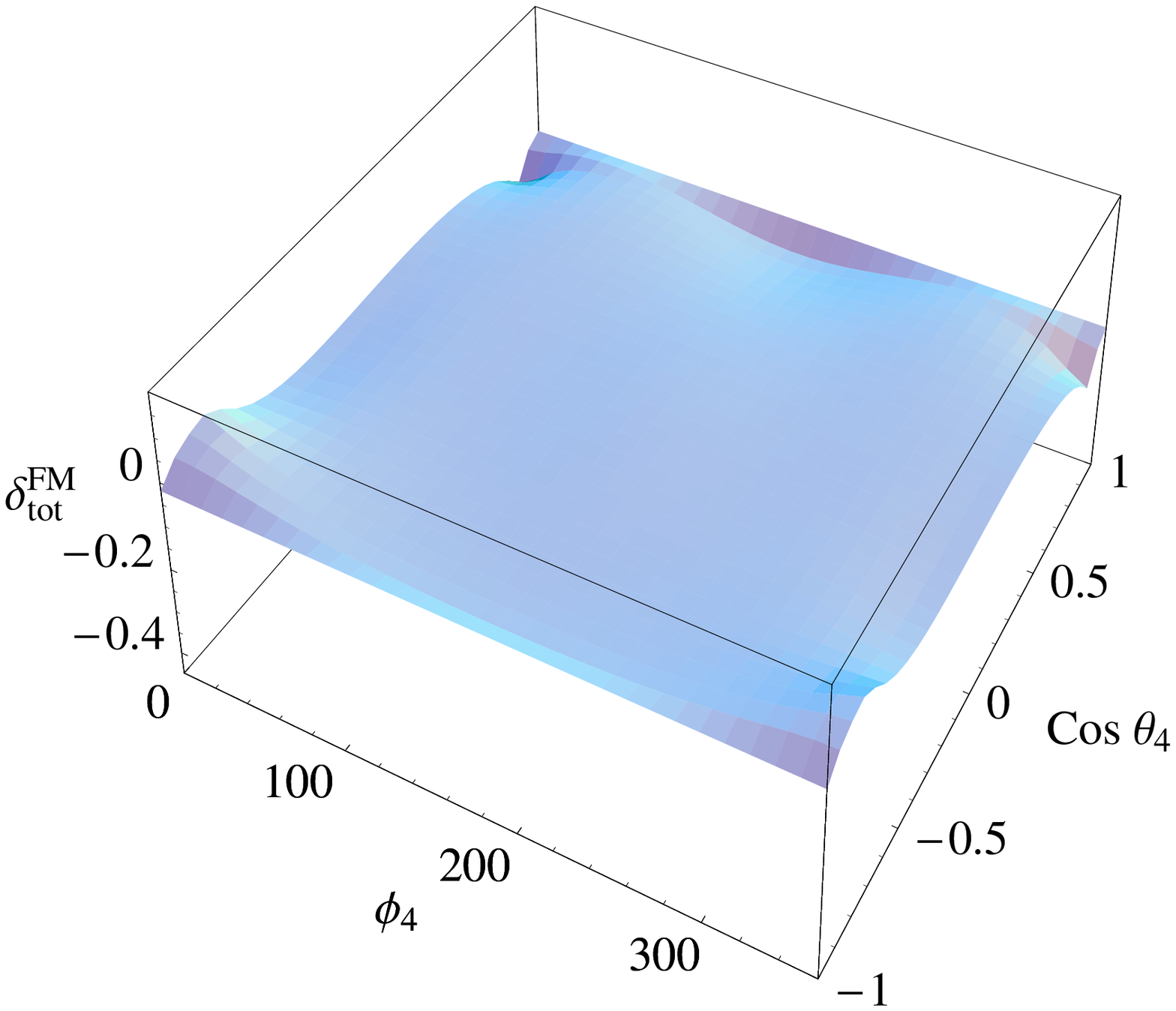}
\par\end{centering}

\caption{Examples of the two-photon box correction angular dependencies for
all three approaches at fixed high momentum transfer, invariant mass
and electron lab energy. All plots are given for $Q^{2}=6.0\, GeV^{2}$,
$W=3.2\, GeV$ and $E_{lab}=9.2\, GeV$.}

\label{Flo:piplus5}
\end{figure}

\begin{figure}
\begin{centering}
\includegraphics[scale=0.34]{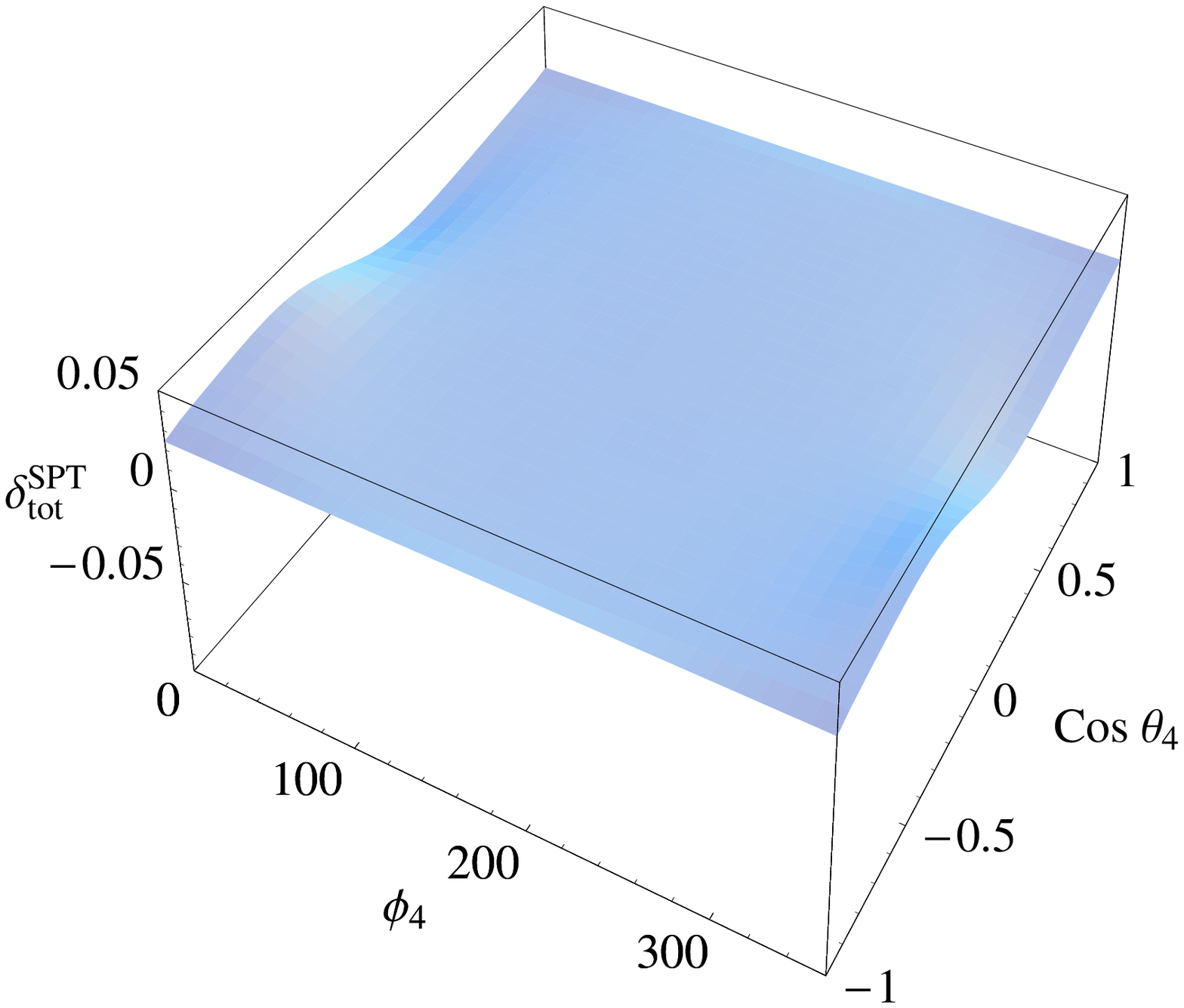}\includegraphics[scale=0.34]{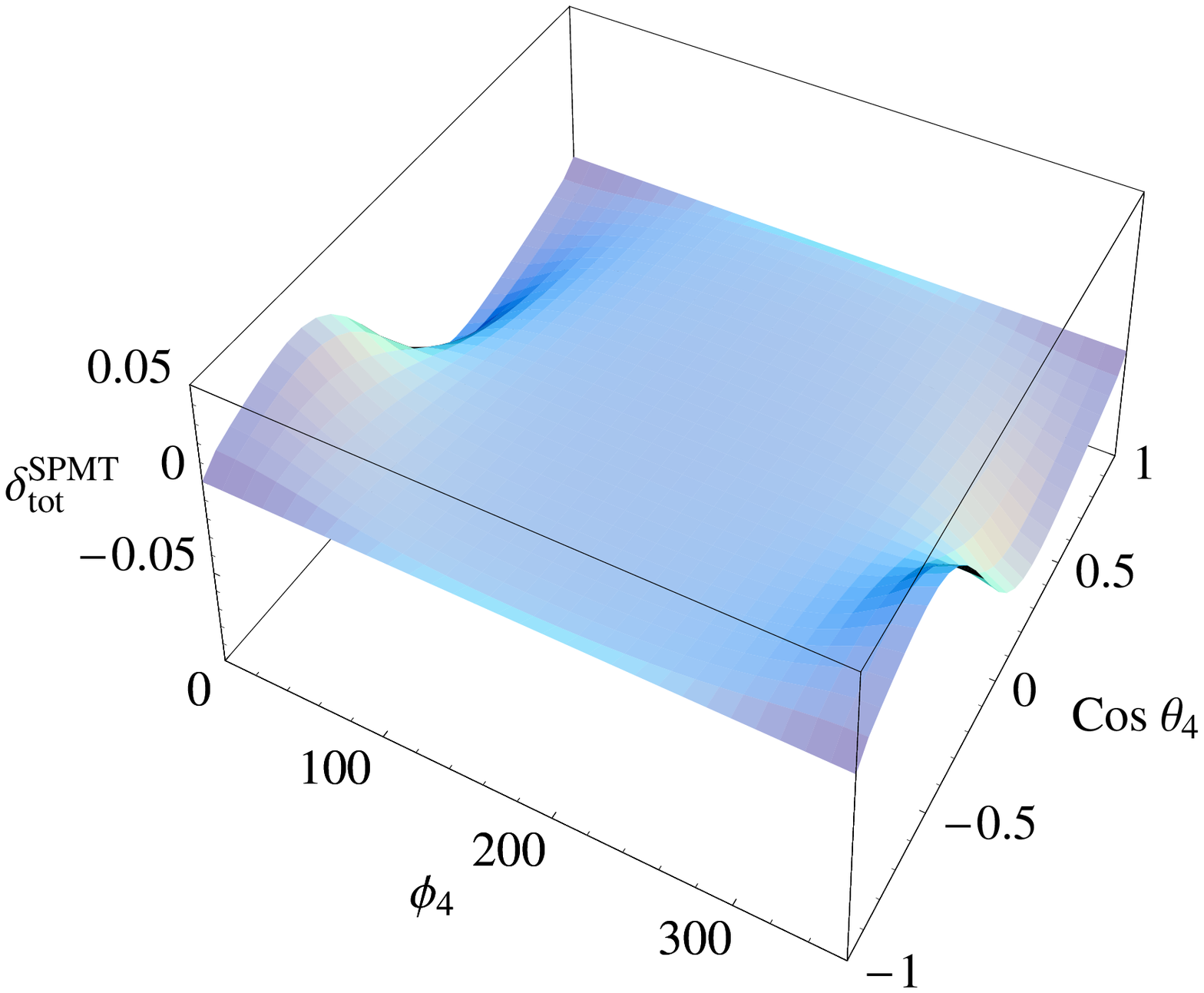}\includegraphics[scale=0.34]{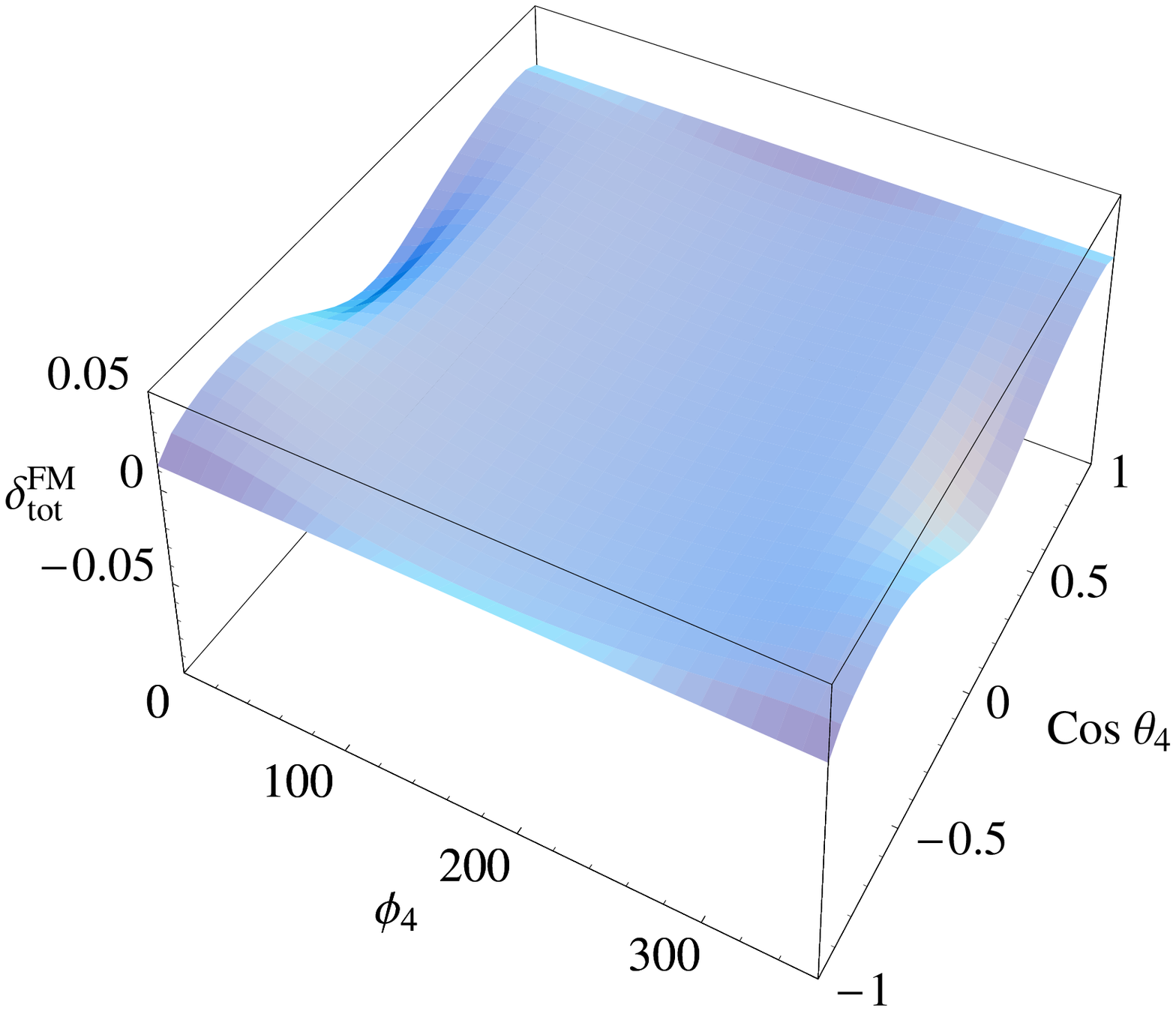}
\par\end{centering}

\caption{Examples of the two-photon box correction angular dependencies for
all three approaches at fixed low momentum transfer, invariant mass
and electron lab energy. All plots are given for $Q^{2}=1.7\, GeV^{2}$,
$W=1.4\, GeV$ and $E_{lab}=5.754\, GeV$.}

\label{piplus5a}
\end{figure}

The exclusive $\pi^{+}$ electroproduction process proved to be a
very successful tool in studies of the transition form factors
to the states in the mass region above $\Delta(1232)$ resonance (see
\citep{CLAS2008}). Since there is a cluster of three states $N(1440)$,
$N(1520)$, $N(1535)$ and at least nine $N^{*}$, $\Delta^{*}$ states
available in the mass region $1.44\, GeV\thicksim1.72\, GeV$, it
is crucial to distinguish them in order to be able to gain information
about the transition form factors. Most of these states are quite
sensitive to both $n\pi^{+}$ and $p\pi^{0}$ channels, but since
the states with isospin 1/2 couple more strongly to the $n\pi^{+}$
than to the $p\pi^{0}$, it is very important to obtain a full picture
about the $\pi^{+}$ electroproduction cross section and extract the unradiated
cross section. Obviously, a full picture requires the inclusion of the radiative corrections. Here, we have extended
calculations of \citep{Afan2002} with the two-photon box correction
for the $\pi^{+}$ electroproduction case. As an example, in the Fig.(\ref{piplus1a})
(left and middle plots), we show the box correction dependencies on
the photon degree of polarization parameter $\epsilon$ for the invariant
mass in the $\Delta$ region ($W=1.232\, GeV$) and for the fixed
high momentum transfers. 
The box correction in the SPT case has a weak dependence on $\epsilon$
and reaches its largest value of $-0.5\%$ for $Q^{2}=7.0\, GeV^{2}$.
On the contrary, the correction calculated with the help of the SPMT
and FM approaches shows a very strong dependence on $\epsilon$, which
is enhanced in the region of the higher momentum transfers. For all  approaches considered here,
the biggest absolute value of the correction is observed for the case
of backward electron scattering. If the invariant
mass is above the mass of $\Delta$ resonance for $Q^{2}=3.0\, GeV^{2}\thicksim7.0\, GeV^{2}$,
we observe that the absolute value of the correction tends to become
smaller in the case of backward electron scattering for the SPMT and
FM approaches only. We find that the correction calculated using the SPT
prescription is not sensitive to the changes of the invariant mass
in the region of $1.232\, GeV\thicksim1.72\, GeV$. For the low momentum
transfer (see Fig.(\ref{piplus1a}) (right plot)), as in the case
of $\pi^{0}$ electroproduction, the absolute value of the correction
obtained in the SPMT and FM approaches decreases substantially, hence
justifying its strong $Q^{2}$ dependence. For the low momentum transfer
of $Q^{2}=0.4\, GeV^{2}$, the dependencies on the invariant mass
$W$, polar and azimuthal angles of the detected $\pi^{+}$ are shown
in Fig.(\ref{Flo:piplus2}). 
In contrast with the high momentum transfers, for $Q^{2}=0.4\, GeV^{2}$
the correction becomes rather small on the absolute scale and all
the approaches agree with each other within $6\%$. We find that the correction
in Fig.(\ref{Flo:piplus2}) (left plot) with fixed low momentum transfer
and $E_{lab}=1.645\, GeV$, is more sensitive on the relative scale
to the variations of the invariant mass in the SPMT and FM approaches  compared to the case of high momentum transfer. For the overall
comparison between corrections obtained at the high and low fixed
momentum transfers, we show dependencies of the two-photon box correction
on the hadronic angular variables in Fig.(\ref{Flo:piplus5}) and
Fig.(\ref{piplus5a}), respectively.

From Fig.(\ref{Flo:piplus5}) and (\ref{piplus5a}) it is evident
that the box radiative correction has a visible effect on the angular
distribution in the SPMT and FM approaches only.

\begin{figure}
\begin{centering}
\includegraphics[scale=0.45]{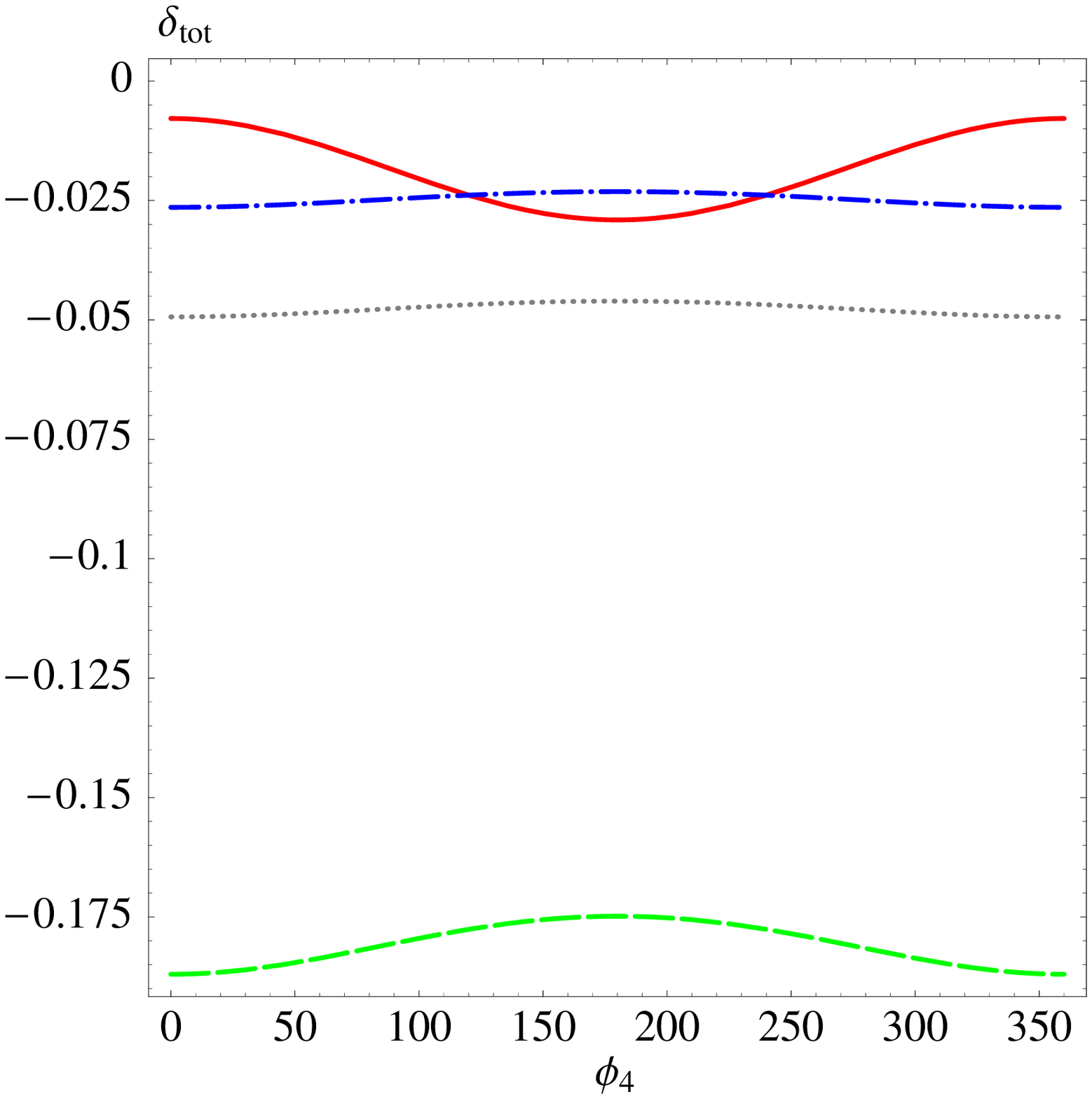}\includegraphics[scale=0.45]{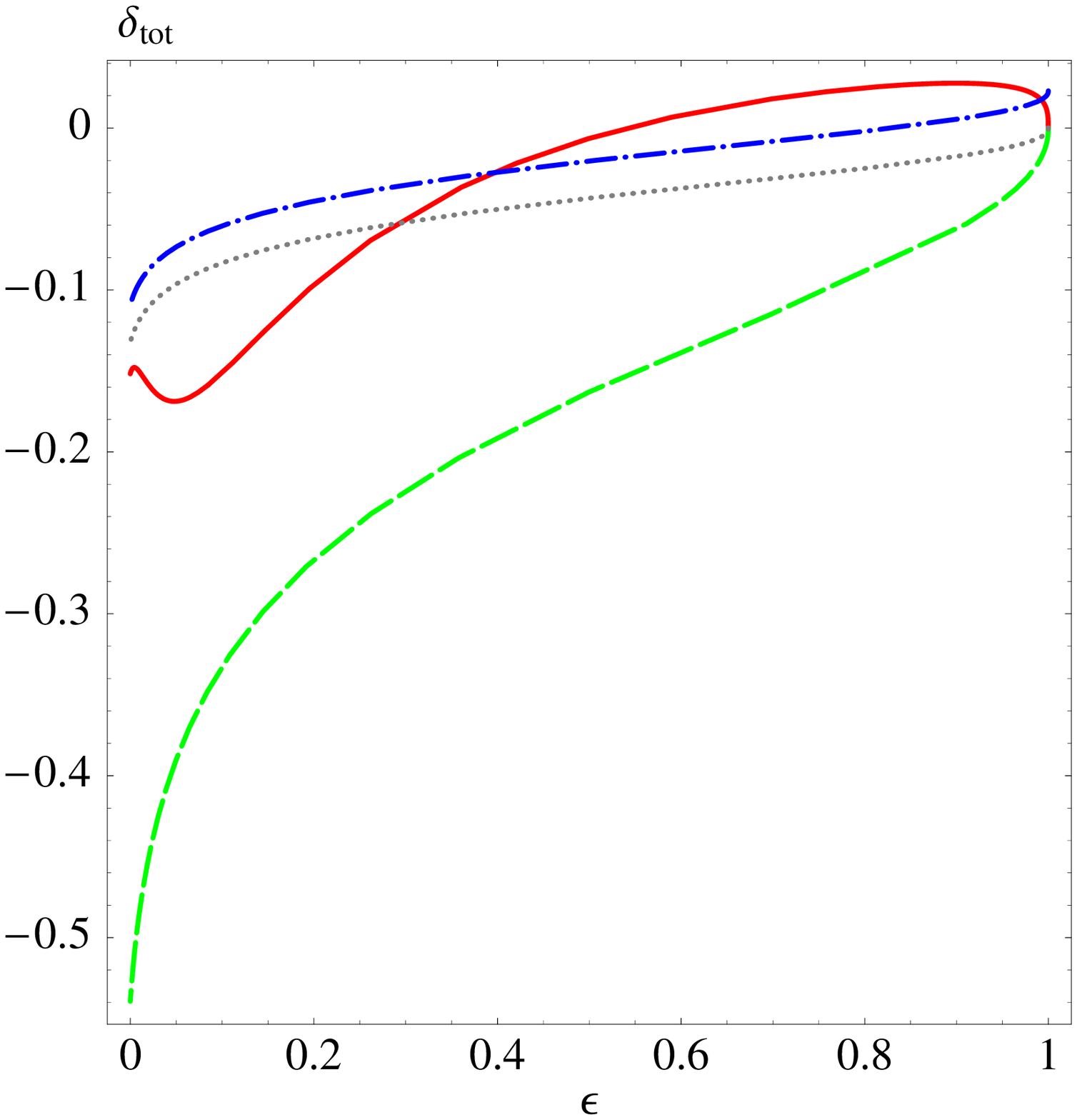}
\par\end{centering}

\caption{$\pi^{+}$ electroproduction azimuthal (left plot, $\theta_{4}\rightarrow0^{\circ}$)
and virtual photon degree of polarization (right plot, $\phi_{4}=90^{\circ}$and
$\theta_{4}\rightarrow0^{\circ}$) dependencies of the two photon
box correction for the fixed $Q^{2}=6.0\, GeV^{2}$, $E_{lab}=10.9\, GeV$
(for the case of azimuthal dependence) and $W=3.2\, GeV$.}

\label{Flo:piplus3}
\end{figure}

Another implementation of the box radiative corrections can be found
in the studies of the $\pi^{+}$ elastic electric form factor. Once
again, in order to extract the Born cross section from the experimental
data, it is essential to apply the radiative corrections to the measured
cross section. For the highest momentum transfer of $Q^{2}=6.0\, GeV^{2}$
and the invariant mass $W=3.2\, GeV$ relevant to the proposed $F\pi$
\citep{Garth2_2008} experiment, Fig.(\ref{Flo:piplus3}) shows the
box correction for the case of forward $\pi^{+}$ electroproduction
($\theta_{4}\rightarrow0^{\circ}$) in the center of mass of final
hadrons.
For all approaches, the azimuthal dependence (see Fig.(\ref{Flo:piplus3})
(left plot)) of the correction is unsubstantial and has a variation
of the order of $2\%$ over the range of $\phi_{4}\in[0,\,2\pi]$. The
range of the correction is in the interval of $-18\%\thicksim-2.5\%$
and the largest absolute value is observed for the correction calculated
in the SPMT approach. Dependencies of the box correction on the degree
of polarization parameter $\epsilon$ are demonstrated in the Fig.(\ref{Flo:piplus3})
(right plot). It is quite interesting to see that the correction obtained
in the SPMT approach exhibits very strong dependence to the variations
of $\epsilon$ for the case of $\pi^{+}$ production in the
forward direction. Overall, the correction calculated in the SPMT
approach shows the largest sensitivity to $\epsilon$ in the forward
direction of the produced charged pion if compared to the SPT and FM approaches.
The fact that the correction calculated using the FM method does not
grow substantially, even for the case of backward scattering ($\epsilon\rightarrow$0),
for the case of deep-virtual kinematics ($Q^{2}=6.0\, GeV^{2},E_{lab}=10.9\, GeV\text{ and }W=3.2\, GeV$)
is most likely evidence of the strong suppression of the two-photon
box amplitude by the monopole form factor in this case. The more detailed
numerical results for the two-photon box correction relevant to the
planned $F\pi$ experiment are given in the Appendix.

\section{Conclusion and Discussion}

In this work, we have evaluated the two-photon exchange correction to pion
electroproduction. Calculations were performed with and without the soft-photon approximation. 
Two versions of the soft-photon approximation were used.
 In the first version, we applied the soft-photon
approximation to the box amplitude consistently in the numerator and
denominator algebra, hence preserving the hadronic current and satisfying
the Ward identity. In the second version, we have followed the proposal
of Maximon and Tjon \citep{Tjon} and applied the soft-photon approximation to the numerator
algebra only. To test the validity of the soft-photon approximation,  we also calculated the two-photon box amplitude
exactly, with no approximations, but instead we had to rely on a model for the reaction dynamics, namely to include the monopole form factor in all hadronic couplings.
 We found that the approximation 
$\mbox{Re}\left[C_{0}\left(\{k_{i},m_{i}\},\{-k_{j},m_{j}\}\right)\right]\thickapprox-C_{0}\left(\{k_{i},m_{i}\},\{k_{j},m_{j}\}\right)$
used in \citep{Tsai} for the three-point scalar integrals resulted
in a systematic shift in the value of correction by $-2.3\%$. After
comparing all three approaches, it became evident that the SPT method 
has produced a very weak kinematic dependence of the correction. Hence,
the soft-photon approximation in the first approach removes the full
kinematic dependence from the correction because the momentum of one of the virtual
photons is neglected in the fermion denominators. This effect is mostly
noticeable at higher momentum transfers. On the contrary, the SPMT 
and FM approaches do conserve the full kinematic dependence which
is enhanced at higher momentum transfers. 
We found that the soft-photon exchange approximation provides a reasonable estimate for the
self-consistent, model-independent calculation, but contributions from hard-photon exchange need to be included.
It is also noted that various recipes for separating the soft-photon exchange contributions may lead
to rather different numerical results.
One may avoid the use of soft-photon approximation at a cost of introducing
models for the two-photon exchange amplitude. Such a model (Formfactor-Model) was used in this work for illustrative purposes.
It is realized that more sophisticated models are required in order to describe the pion electroproduction in the wide range of kinematics considered here.

In conclusion, we found that the two-photon exchange contributions to the pion electroproduction may play a significant role
for high momentum transfers and need to be included in the data analysis of precision experiments in electron scattering.

\begin{acknowledgments}
The authors are grateful to V. Kubarovsky and G. M. Huber for explaining the relevant experiments. A.Al. and S.B. thank the Theory Center at Jefferson
Lab for their hospitality in 2011 and 2012. A.Af. thanks L. Maximon for useful discussions.
This work was partially supported by the Natural
Sciences and Engineering Research Council of Canada.

\end{acknowledgments}

\section{Appendix}

Numerical results for the two-photon box correction for the planned
$F\pi$ experiment are demonstrated in the table Tbl.\ref{tbl1} for
the $\theta_{4}\rightarrow0^{\circ}$.
\begin{table*}
\begin{centering}
\begin{tabular}{|c|c|c|c|c|c|}
\hline 
$E_{e}$(GeV) & $\epsilon$ & $\delta_{tot}^{SPT}$ & $\delta_{tot}^{SPT}-\alpha\cdot\pi$ & $\delta_{tot}^{SPMT}$ & $\delta_{tot}^{FM}$\tabularnewline[2mm]
\hline 
\noalign{\vskip1mm}
\hline 
\multicolumn{6}{|c|}{$Q^{2}=0.3\, GeV^{2},$ $W=2.20\, GeV$, }\tabularnewline
\hline 
\hline 
2.80 & 0.341 & -0.0171 & -0.0400 & -0.1494 & -0.1338\tabularnewline
\hline 
3.70 & 0.657 & -0.0027 & -0.0257 & -0.0920 & -0.0786\tabularnewline
\hline 
4.20 & 0.747 & 0.0013 & -0.0216 & -0.0769 & -0.0634\tabularnewline
\hline 
\hline 
\multicolumn{6}{|c|}{$Q^{2}=1.6\, GeV^{2},$ $W=3.00\, GeV$}\tabularnewline
\hline 
\hline 
6.60 & 0.387 & -0.0271 & -0.0501 & -0.1810 & -0.1394\tabularnewline
\hline 
8.80 & 0.689 & -0.0084 & -0.0314 & -0.1096 & -0.0593\tabularnewline
\hline 
9.90 & 0.765 & -0.0038 & -0.0268 & -0.0927 & -0.0431\tabularnewline
\hline 
\hline 
\multicolumn{6}{|c|}{$Q^{2}=2.45\, GeV^{2},$ $W=3.20\, GeV$}\tabularnewline
\hline 
\hline 
7.40 & 0.265 & -0.0393 & -0.0623 & -0.2293 & -0.1696\tabularnewline
\hline 
8.80 & 0.505 & -0.0209 & -0.0439 & -0.1567 & -0.0842\tabularnewline
\hline 
9.90 & 0.625 & -0.0134 & -0.0363 & -0.1281 & -0.0547\tabularnewline
\hline 
10.90 & 0.702 & -0.0087 & -0.0316 & -0.1106 & -0.0387\tabularnewline
\hline 
\hline 
\multicolumn{6}{|c|}{$Q^{2}=3.5\, GeV^{2},$ $W=3.10\, GeV$}\tabularnewline
\hline 
\hline 
7.90 & 0.304 & -0.0338 & -0.0568 & -0.2159 & -0.1111\tabularnewline
\hline 
9.90 & 0.587 & -0.0145 & -0.0375 & -0.1377 & -0.0341\tabularnewline
\hline 
10.90 & 0.671 & -0.0095 & -0.0325 & -0.1182 & -0.0193\tabularnewline
\hline 
\hline 
\multicolumn{6}{|c|}{$Q^{2}=4.50\, GeV^{2},$ $W=3.28\, GeV$}\tabularnewline
\hline 
\hline 
8.80 & 0.220 & -0.0444 & -0.0673 & -0.2566 & -0.1307\tabularnewline
\hline 
9.90 & 0.400 & -0.0283 & -0.0512 & -0.1905 & -0.0608\tabularnewline
\hline 
10.90 & 0.520 & -0.0201 & -0.0430 & -0.1577 & -0.0315\tabularnewline
\hline 
\hline 
\multicolumn{6}{|c|}{$Q^{2}=5.25\, GeV^{2},$ $W=3.20\, GeV$}\tabularnewline
\hline 
\hline 
8.80 & 0.188 & -0.0468 & -0.0697 & -0.2722 & -0.1203\tabularnewline
\hline 
9.90 & 0.373 & -0.0387 & -0.0616 & -0.2376 & -0.0479\tabularnewline
\hline 
10.90 & 0.498 & -0.0206 & -0.0436 & -0.1631 & -0.0190\tabularnewline
\hline 
\hline 
\multicolumn{6}{|c|}{$Q^{2}=6.00\, GeV^{2},$ $W=3.20\, GeV$}\tabularnewline
\hline 
\hline 
9.20 & 0.177 & -0.0482 & -0.0711 & -0.2798 & -0.1085\tabularnewline
\hline 
9.90 & 0.298 & -0.0353 & -0.0583 & -0.2250 & -0.0558\tabularnewline
\hline 
10.90 & 0.435 & -0.0247 & -0.0477 & -0.1809 & -0.0185\tabularnewline
\hline 
\end{tabular}
\par\end{centering}

\caption{Two-photon $\pi^{+}$ electroproduction box correction for the proposed
$F\pi$ experiment with the $\phi_{4}=90^{\circ}$ and $\theta_{4}\rightarrow0^{\circ}$.}
\label{tbl1}
\end{table*}


\begin{thebibliography}{References}






\bibitem{12GeV} http://www.jlab.org/12GeV/.

\bibitem{TPE} S.~J.~Brodsky, C.~E.~Carlson, Y.~-C.~Chen and M.~Vanderhaeghen,
  Phys.\ Rev.\ D {\bf 72}, 013008 (2005); P.~G.~Blunden, W.~Melnitchouk and J.~A.~Tjon,
  Phys.\ Rev.\ C {\bf 72}, 034612 (2005).
  
 \bibitem{CLAS2008}K. Park et al. [CLAS Collaboration], Phys. Rev. C
{\bf 77}, 015208 (2008).

\bibitem{Garth1_2008}H. P. Blok et al. Jefferson Lab $F_\pi$ Rev. C {\bf 78}, 045202 (2008).

\bibitem{Garth2_2008}G. M. Huber et al. Jefferson Lab $F_\pi$ Rev. C {\bf 78}, 045203 (2008).

\bibitem{E07007}C. Camacho et. al. and Jefferson Lab Hall A Collaboration,
Jefferson Lab PAC 31 Proposal, http://wwwold.jlab.org/exp\_prog/proposals/07/PR-07-007.pdf
(2006).

\bibitem{P1206101}G.M. Huber et. al., Jefferson Lab PAC 30 Proposal,
http://wwwold.jlab.org/exp\_prog/proposals/06/PR12-06-101.pdf

\bibitem{Ji97}	X. Ji, Phys. Rev. D {\bf 55} (1997) 7114;
		A.V. Radyushkin, Phys. Rev. D {\bf 56} (1997) 5524.

\bibitem{Col97}	J.C. Collins, L. Frankfurt, and M. Strikman, 
		Phys. Rev. D {\bf 56} (1997) 2982-3006.


\bibitem{Mo&Tsai}L. W. Mo and Y. S. Tsai, Rev. Mod. Phys. {\bf 41}, 205 (1969);
Y.S. Tsai, Preprint SLAC, PUB - 848, (1971).

\bibitem{Haprad} I. Akushevich, N. Shumeiko, A. Soroko, Eur. Phys. J. C {\bf 10}, 681 (1999).

\bibitem{Diffrad} I. Akushevich, Eur. Phys. J. C {\bf 8}, 457 (1999).

\bibitem{Afan2002} A.~Afanasev, I.~Akushevich, V.~Burkert and K.~Joo, Phys. Rev. D {\bf 66}, 074004 (2002).


\bibitem{Bar77}D. Y. Bardin and N. M. Shumeiko, Nucl. Phys. B {\bf 127},
242 (1977).



\bibitem{Blunden_pi_2010}P. G. Blunden, W. Melnitchouk, J. A. Tjon,
Phys. Rev. C {\ bf 81}, 018202 (2010). 

\bibitem{Blunden_ep_2005}P.G. Blunden,W. Melnitchouk, and J. A. Tjon,
Phys. Rev. C {\bf 72}, 034612 (2005).
 
\bibitem{Dong_and_Wong_2010}Y.-B.~Dong, S.D.~Wang, Phys. Lett. B {\bf 684},
123 (2010).

\bibitem{Borisyuk_pi_2010}D. Borisyuk, A. Kobushkin, Phys.\ Rev.\ C {\bf 83}, 025203 (2011).

\bibitem{LoopTools}T. Hahn and M. Perez-Victoria, Comput. Phys. Commun.
118, 153 (1999).


\bibitem{Tsai}Y.S. Tsai, Phys. Rev. {\bf 122}, 1898 (1961).

\bibitem{Tjon}L.C. Maximon, J.A. Tjon, Phys.Rev. C {\bf 62}, 054320 (2000). 

\bibitem{Kuraev2006}E. A. Kuraev {\it et al.}, Phys. Rev. D {\bf 74}, 013003
(2006).

\bibitem{Manohar} A. Manohar, arXiv:hep-ph/9305298 (1993).


\bibitem{QEDRadCor}A. Afansev, I. Akushevich, V. Burkert, K. Joo,
Phys. Rev. D 66, (2002) 074004 

\bibitem{CLAS}A. N. Villano et al., Phys. Rev. C80 (2009) 035203


\bibitem{CLASa}K. Joo et al. (CLAS), Phys. Rev. Lett. 88, 122001,
(2002).

\bibitem{CLASb}K. Joo et al. (CLAS), Phys. Rev. C 68, 032201 (2003).

\bibitem{CLASc}A. Biselli et al. (CLAS), Phys. Rev. C 68, 035201
(2003).

\bibitem{CLASd}K. Joo et al. (CLAS), Phys. Rev. C 70, 042201 (2004).













\end{thebibliography}
\end{document}